\documentclass[rmp,10pt]{revtex4}
\usepackage[utf8]{inputenc}
\usepackage{natbib}
\setcitestyle{numbers}
\usepackage{mathtools}
\usepackage{amsmath}
\usepackage{amsfonts}
\usepackage{amssymb}
\usepackage{epsfig}
\usepackage{wrapfig}
\usepackage{changepage}
\usepackage{float}
\usepackage{wrapfig}

\newcommand{\beq}{\begin{equation}}
\newcommand{\eeq}{\end{equation}}

\begin{document}

\title{Generalized Boltzmann distribution for systems out of equilibrium}

\author{Milo M. Lin}
\affiliation{Green Center for Systems Biology, Department of Biophysics, and the Center for Alzheimer's and Neurodegenerative Diseases, University of Texas Southwestern Medical Center, Dallas, TX 75390 \\
Milo.Lin@UTSouthwestern.edu}

\begin{abstract}
     The Boltzmann distribution predicts the collective behavior of systems at thermodynamic equilibrium as a function of their constituent parts. Yet most systems in nature are not at equilibrium, and a unified theory of their behavior does not currently exist. Here, I show that the Boltzmann distribution is a special case of a general distribution that governs all stochastic systems, even if far from equilibrium. The generalized Boltzmann distribution is explained as an analog of the voltage equation in electronics, where resistors, batteries, node voltages, and path currents correspond to equilibrium rate constants, driven rate constants, probabilities, and probability flows, respectively. The general distribution recapitulates known properties of weakly driven systems and enables new closed-form solutions for strongly-driven systems. These solutions provide insight into fundamental limits on system performance, and experimental data show that living systems can operate at those limits. The formal mapping between non-equilibrium systems and electronic circuits may provide a unified framework to simplify, understand, and ultimately control the behavior of complex non-equilibrium systems. 
\end{abstract}

\maketitle

\section{Introduction}

A system at thermodynamic equilibrium obeys detailed balance, meaning that the average transition rate from any state of the system to any other state is exactly balanced by the reverse rate. The probability  $P_i$ of finding the system in state $i$, relative to any other state $j$, is given by the Boltzmann distribution \cite{boltzmann}:
\begin{equation}
P_i \mathrm{e}^{\beta G_i}=P_j \mathrm{e}^{\beta G_j},
\end{equation}
where $G_i$ is the free energy of state $i$ and $\beta=1/k_BT$ is the reciprocal of Boltzmann's constant times the temperature. Eq. 1 provides a direct relationship between local parameters of individual states (free energies) and the probabilities of those states to occur, even though the probabilities arise from collective global dynamics between all the states. This property enables prediction of equilibrium behavior for systems with many possible states, for which computing transitions between all states would be intractable. Consequently, the Boltzmann distribution is the cornerstone of statistical mechanics, being used to explain a wide variety of collective phenomena from the molecular to the macroscopic. 

However, the Boltzmann distribution does not apply to systems out of equilibrium. Such systems do not obey detailed balance because they can take in net energy from their environments in order to sustain net flow of probability between states, thereby performing work and dissipating heat. Most systems found in nature fall into this category \cite{kinesin_efficiency, f1atpase, rapoport_pump}. In addition to doing work, energy expenditure is often necessary in order to endow complex systems like cells with characteristics--such as all-or-none regulation--that would be impossible at equilibrium \cite{gunawardena}. More generally, processes with non-thermal sources of stochasticity can be modelled as nonequilibrium systems. Examples include turbulent fluid flows\cite{ruelle}, genome evolution under natural selection \cite{fisher}, and economic markets \cite{ingber}. Most of these systems are driven very far from equilibrium, such that even their qualitative behavior deviates from equilibrium theory. Despite their ubiquity, nonequilibrium systems remain poorly understood. Theoretical characterization has focused on special cases in which systems are driven slightly away from equilibrium \cite{prigogine, Jaynes, dill1, dill2} or transitions between states of equilibrium \cite{Jarzynski}. Far from equilibrium, most of these theories do not apply \cite{martyushev}. The few that do, such as the fluctuation theorems \cite{evans, Gallavotti, Crooks} or dissipation-fluctuation bounds \cite{seifert_efficiency1, todd}, do not predict collective behavior (such as $P_i$) as a function of the constituent interactions, or do so by combinatorially enumerating paths between all possible states \cite{hill, schnakenberg, zia}. The key barrier has been the lack of a relationship, akin to the Boltzmann distribution, that explains collective behaviors in terms of local parameters. Consequently, there are few organizational or design principles for nonequilibrium systems.
\par
Here, I formulate a theory that generalizes the Boltzmann distribution to apply to any memory-less dynamical system, and use the theory to derive new emergent properties of nonequilibrium systems. Like the Boltzmann distribution, the new theory links individual state parameters and collective behaviors, but is generally applicable to systems that consume and dissipate energy and mass, have nonlinear rate coefficients, and may be time-varying. This generalization arises from an exact mapping from any Markov process to a circuit in which nodes are the states and currents are probability fluxes between states which are driven by differences in a "probability potential" in the same way that currents are driven by an electrostatic potential. In this mapping, there are only three kinds of elementary components---"resistors," "grounds", and "batteries"---which correspond to the undriven processes, mass sources, and energy sources, respectively. A key difference from conventional linear electronic circuits is that the "batteries" in a probability circuit have built-in feedback, providing for special properties. However, like electronic circuits, the resistors in a probability circuit can be lumped together, simplifying a complex system into a representation consisting of only the irreducible collective variables.  
\par
This mapping allows circuit theorems to be applied to nonequilibrium statistical mechanics, from which they take on new meaning. Here, I show that they enable closed-form solutions of classic nonequilibrium processes, including signaling/catalysis, self-assembly, and kinetic proofreading. These solutions expose simple universal limits and tradeoffs governing such processes. More generally, the theory leads to two laws governing maximum energy efficiency and reciprocal response to driving. These results initiate a unifying framework in which the global emergent behaviors of non-equilibrium systems can be understood without exhaustive enumeration of their details.

%------------------------------------------------

\section{The Probability Flow Equation}

Consider a system that can exist in many possible states with rate coefficients between the states such that the system will reach equilibrium. We define two states  $m$ and $n$ to be "adjacent" if there exists a pathway between them that does not traverse any other states. Such a system is completely characterized by forward rate constants between neighboring states $k_{mn}$ and their free energies $G_m$, with the $k$'s representing the usual equilibrium rate constants. If we now drive the system away from equilibrium by adding a driven rate constant $\alpha_{mn}$ to $k_{mn}$, we will observe a net probability flow, denoted by the current $I_{mn}$, from state $m$ to $n$ that is the difference between the forward and reverse rates: $I_{mn} = P_m(k_{mn}+\alpha_{mn})-P_n k_{nm}$. Let us now define a probability "potential" that drives such a flow:
 \begin{equation}
V_{m} \equiv P_{m} \mathrm{e}^{\beta G_m}.
\end{equation} 
Intuitively, the potential of a state is (up to normalization by the partition function) its probability divided by its equilibrium probability: differences in the potential provide the driving force for probability flows. If there is a mass source or sink in the system, which is a state whose probability (i.e. potential) remains unchanged by probability flow into or out of the state, such states correspond to "grounds."
The relation between the probability flow, $I$, and our definition of the probability potential, $V$, can be established by first defining 
the "resistance," $R_{mn}$, between states $m$ and $n$ (Fig. 1):
 \begin{equation}
R_{mn} \equiv {{\mathrm{e}^{\beta G_m}} \over {k_{mn}}}=R_{nm}.
\end{equation} 

Note that the second equality in Eq. 3 follows because the equilibrium forward transmittance is equal to the equilibrium backward transmittance; hence the resistance is directionally symmetric, just like the behavior of standard resistors in electronic circuits. This property will be crucial for the simplification of complex systems. 
 Finally, define the "battery" driving transitions from $m$ to $n$ as:
 \begin{equation}
\mathcal{E}_{mn} \equiv {\alpha_{mn} \over {k_{mn}}} \mathrm{e}^{\beta G_m}P_{m},
\end{equation} 
which is proportional to the driven rate constant divided by the equilibrium rate constant and is thus zero when the transition between $m$ and $n$ is not driven. $\mathcal{E}_{mn}$ is also proportional to the potential at $m$ as defined by Eq. 2; the battery is therefore a potential-feedback battery. 
\par
Using the definitions from Eqs.2-4, we can obtain (See Supp. materials for all derivations) the probability flow between any two states $i$ and $j$ in terms of the difference in probability potential between the states: 
\begin{equation}
V_{j} -V_{i} = {\sum\limits_{m=i}^{n=j}(\mathcal{E}_{mn}-R_{mn}I_{mn})},
\end{equation}
 where $m$ and $n$ denote the intermediate adjacent states on \textit{any} the path from $i$ to $j$ (See, for example, two alternate paths in Fig. 2A). Fig. 1 summarizes the translation of local dynamical relationships in state space to circuit elements. 
 \par
 Electronic circuits have been previously used as analogies of dynamical systems before \cite{zia, Oster, baez}, but Eqs. 2-5 establishes a rigorous mathematical mapping and exposes the essential similarities and differences. For example, the feedback nature of "batteries" in a probability circuit sets them apart from their counterparts in electronic circuits, with the feedback qualitatively differentiating near-equilibrium linear-response behavior from far-from-equilibrium nonlinear behavior. But like in electronic circuits, the "resistors" in the probability circuit can be systematically lumped together using the usual rules for combining resistors in parallel or in series (Fig. 2B), or using the star-mesh transform \cite{EEbook}. Using these rules, any complex system can be simplified to a minimal set of irreducible collective variables that are explicit functions of the microscopic equilibrium parameters. The reducibility of any system is therefore dictated by the number and placement of driven transitions (the "batteries") within the state space. Seemingly different or unrelated dynamical systems may reduce to the same equivalent circuit, providing a way to categorize dynamical systems according to their irreducible circuit representations.
 \par
 The probability flow equation (PFE, Eq. 5) is general; it holds for time-dependent behavior as well as for steady state, and applies to systems relaxing to equilibrium or driven systems undergoing nonlinear dynamics. In the special case of equilibrium, there are no batteries because $\alpha_{mn} = 0$ for all $m$ and $n$. In this case, all currents are zero, all states are "equipotential," the system is at equilibrium, and the PFE (Eq. 5) reduces to the Boltzmann distribution Eq. 1.

\section{Applications: design principles and limits of biomolecular processes}
Well-known biomolecular processes provide good examples to illustrate the value of the PFE and to compare its predictions with experimental data. In principle, the PFE should unify diverse problems that have been studied using case-specific formalisms, with well-known equations emerging as special cases of Eq. 5. More importantly, the circuit formulation should lead to new predictions in well-studied systems; for example, providing fundamental limits on the capabilities of arbitrarily complex systems. These limits specify tradeoffs between enhancing/suppressing occupancy of states, currents, and energy costs. Experimental data are used to quantitatively show that living systems obey--and are often performing at the edge of--these limits. 

\subsection{Catalysis and signaling}
The class of systems whose state space consists of $N$ states arranged in a single loop (Fig. 3a, center) encompasses molecular machines that perform cyclic tasks, such as enzymes and signaling molecules (Fig. 3a, left). When mapped to a probability flow circuit, sequential undriven steps along the loop correspond to resistors in series. The total resistance of any portion of the loop from state $i$ to $j$ is the sum of the individual resistances in the portion: $R_{i;j} \equiv \sum\limits_{m=i}^{j} R_{m m+1}$. The total resistance of the loop is: $R_{\mathrm{tot}} \equiv R_{1;N}$ (note that $N+1 = 1$ due to periodicity).  
At steady-state, the current must be the same everywhere along the loop: $I \equiv I_{mn}$. Solving Eq. 5 with the condition that probabilities must add to one, the current and probabilities are given by:
\begin{subequations}
\begin{equation}
I =  {{1}\over{Z}} { { {{\alpha} \over {k_{12}}} } \over {R_{\mathrm{tot}}+{{\alpha} \over {k_{12}}} \sum\limits_{i=2}^N R_{i;N} P_i^*}}.
\end{equation}
\begin{equation}
P_{i}= P_{i}^*\Bigg({{R_{\mathrm{tot}}+(1-\delta_{i1}){{\alpha} \over {k_{12}}}R_{i;N}} \over {R_{\mathrm{tot}}+{{\alpha} \over {k_{12}}} \sum\limits_{j=2}^N R_{j;N} P_j^*}}\Bigg)
\end{equation}

\end{subequations}
$P_{i}^*$ is the equilibrium probability of state $i$, the Kronecker delta $\delta_{ij} = 1$ if $i=j$ and 0 otherwise, and the partition function $Z$ is the normalization constant of the undriven system. 
As an instance of Eq. 6a, consider a simple system consisting of three states ($N=3$), which represents the canonical simple model of enzymatic reactions. Enzymes transform substrate into products; the concentrations of these three species are denoted [Eo], [S], and [P], respectively, with the three states of the total system shown in Fig3A, right. The current flow in the three-state loop is the rate at which substrate is transformed into product. The system is driven from equilibrium by the continual extraction of product and replenishment of substrate (red arrow in the state space, the battery in the probability circuit). Substrate binding is proportional to [S] : $k_{23}\equiv k_f [S]$ and $k_{31} \equiv k_{\mathrm{cat}}$ is the enzyme catalytic rate. The current, Eq. 6a, rewritten in terms of these variables is: 
 \begin{equation}
v ={ {k_{\mathrm{cat}}[\mathrm{E_o}][\mathrm{S}]} \over {K_{\mathrm{M}}+[\mathrm{S}]}} \Bigg({{\alpha P_1^*} \over {\alpha P_1^*+ { {k_{\mathrm{cat}}[\mathrm{S}]} \over  {K_{\mathrm{M}}+[\mathrm{S}]}}}} \Bigg) ,
\end{equation}
where the Michaelis constant $K_{\mathrm{M}}\equiv (k_{\mathrm{cat}}+k_r)/k_f$, and $v=[\mathrm{Eo}]I$ is the number current rather than the probability current. If $\alpha$ is much larger than the equilibrium rate of conversion from substrate back to product, the rate expression converges to the well-known Michaelis-Menton relation for irreversible reactions  \cite{michaelis}. Thus, Eq. 7 is a generalized Michaelis-Menton relation applicable to reversible enzymes.
\par

 Eq. 6b directly shows how nonequilibrium systems can selectively promote and stabilize energetically unfavorable states, a feature unavailable to equilibrium systems. The ratio in Eq. 6b is the amplification of the equilibrium probability, which is determined by the relative magnitudes of the driven versus non-driven rate coefficients (${{\alpha} \over {k_{12}}}$). Consequently, the maximum amplification of any state is: 
\begin{equation}
\Big| \ln\Big({{P_i}\over{P_i^*}}\Big) \Big| < {{\mathcal{P}}\over{I}}\cdot {{1}\over{k_{\mathrm{B}}T}},
\end{equation}
where $\mathcal{P}$ is the power consumption (energy cost per time; see Section IV.A and Supp. materials). The right-hand-side of Eq. 8 is thus the energy cost per cycle in units of $k_\mathrm{B}T$. For any cyclic system, \textit{no} state can be amplified or suppressed by a factor more than the exponential of the energy cost per cycle. Eq. 8 is a tight bound. As an example, consider the five-state model of Ras, the master signalling G-protein (Fig. 3A, left). Computing the probability amplification using the measured rate constants in the "inactive" versus "active" conditions (shown in black and white, respectively, depending on the presence of the GDP release factor Sos) \cite{ras,ras2sos} shows that some states of this system are close to the amplification limit, which is set by the energy released upon hydrolyzing GTP (Fig. 3A, right).
\par

\subsection{Kinetic proofreading}
The circuit diagram for kinetic proofreading, the general mechanism of biomolecular error correction, is shown in Fig. 3B, Center. A specific example is the translation of the genetic code into proteins (Fig.3B,left). The correct and erroneous tRNA match can both be bound, with the erroneous binding intermediate being $\Delta$ less energetically stable than the correct match. Following binding, the complex is driven to the activated state via binding of GTP, corresponding to the potential drop $\mathcal{E}$, after which protein elongation can proceed to completion or fail. Upon successful completion (whether correct or erroneous), the process is then reset back to the origin state using $\mathcal{E}_{\mathrm{reset}}$, corresponding to the maintenance of fixed product and reactant concentrations. The minimum error was shown previously to be $e^{-2\beta \Delta}$ \cite{hopfield}, and the limiting behavior of error and speed in different regimes was derived for generalized proofreading schemes \cite{murugan}. The lower bound on the power required for a given speed has been shown to decrease as the logarithm of the error \cite{gunawardena_proofreading}. However, in the biologically-relevant regime, numerical simulations show that the power requirements far exceed this lower bound \cite{gunawardena_proofreading}.  In fact, experiments show that the catalytic efficiency (speed divided by power) increases linearly with the error \cite{linear_tradeoff}. 

\par
The probability flow equation reconciles these results, showing that they arise from a direct tradeoff between error, speed, and power. To explain, in terms of the circuit variables (Fig. 3B, center), the speed $=I_2$, error $\epsilon=I_4/I_2$, and the dissipation rate (power consumption) $\mathcal{P} = k_{\mathrm{B}}T\ln(1+{{\alpha}\over{k}})(I_1+I_2+I_3+I_4)$. The steady-state voltage equations taken over the four inner loops of the circuit can be solved analytically to give the error as a function of the power, speed, and resistors (See Supp. material). In the near-equilibrium regime ($\alpha << k$), the expression can be simplified to give: $\epsilon > e^{-\beta \mathcal{P}/I_2}e^{-\beta \Delta}$, which is the previously-known lower bound \cite{gunawardena_proofreading}. Thus, this trade-off bound is only a tight constraint on the system behavior if the proofreading capacity is close to equilibrium, explaining the difference between previous theoretical and simulation results \cite{gunawardena_proofreading}. In the biologically-applicable far-from-equilibrium regime ($\alpha >> k$), the expression simplifies to a new relation describing the fundamental trade-off between error $\epsilon$ and energy cost per cycle $\mathcal{P}/I_2$:  
 \begin{equation}
\epsilon > \Bigg({{\mathcal{P}/I_2}\over{\mathcal{P}/I_2-k_{\mathrm{B}}T\ln(1+{{\alpha}\over{k}})}}\Bigg) e^{{-2\Delta}\over{k_{\mathrm{B}}T}},
\end{equation}
  The only parameters that influence the trade-off are $\Delta$, $T$, and $\alpha/k$. Eq. 9 predicts that the accuracy (reciprocal of the error) decreases linearly with the catalytic efficiency, explaining the experimental data \cite{linear_tradeoff}. Fig. 3B,right shows that, for the correct translation of the codon AAA over the erroneous AAU, the protein elongation machinery achieves the optimal bound allowed by Eq. 9 even as it trades accuracy for catalytic efficiency under different magnesium ion concentrations; see white circles in Fig. 3B, right. The system is tuned to the optimal bound of Eq. 9 by decreasing $R_1/R_2$ (Fig. 3B, right); this can be done, for example, by tuning (i.e. mutating) the undriven rate constants of the codon binding step relative to the hydrolysis step. 

\par

\subsection{Microtubule self-assembly}
The elementary steps constituting the dynamic instability of microtubule self-assembly is shown in Fig.3C,left \cite{dynamic_instability}, a process that contains an infinite number of possible states in the thermodynamic limit. The reversible assembly of tubulin dimers, with equilibrium (GDP-bound) forward and reverse rate constants $k_f$ and $k_r$, is driven by tighter binding of GTP-bound tubulin dimers with a free energy difference of $G$; thus, $k_f/k_r=e^{-\beta G}$. Here, the ratio of GTP to GDP concentration provides the drive to push the system out of equilibrium, with the driven forward rate constant, $\alpha$, being proportional to [GTP] up to a saturation concentration (See Supp. material). Assembly is counteracted by the rate of catastrophe $f_{\mathrm{cat}}$, which leads to complete dis-assembly of the microtubule in a manner that depends on both regulatory molecules and [GTP] \cite{luke2}. In the absence of rescue from catastrophe, the microtubule length probability distribution $P(L)$ reaches a bounded steady state with well-known mean length $<L> = (\alpha+k_{\mathrm{f}}-k_{\mathrm{r}})/f_{\mathrm{cat}}$ when driven far from equilibrium \cite{microtubule_leibler}. In this regime, changing $f_{\mathrm{cat}}$ alone leads to a proportional change in the mean length, permitting regulatory molecules acting on the ends of growing microtubules to tune microtubule length in a sub-stoichiometric manner. Such a mechanism occurs during the eukaryotic cell cycle, where changes in catastrophe rates alone produce the decrease in microtubule length necessary for cell division \cite{mitchison}. 

\par 

To understand how regulation of catastrophe depends on nonequilibrium driving, I mapped this process to the probability flow circuit formalism (Fig. 3C,center), and solved Eq. 5 to obtain the closed-form expression for the steady state length distribution of microtubules:

\begin{equation}
P(L)= {{f_{\mathrm{cat}}}\over{f_{\mathrm{cat}}-\alpha (e^{\beta G}-1)}}P_1e^{-\beta G(L-1)}+{{\alpha(e^{\beta G}-1)}\over{\alpha (e^{\beta G}-1)-f_{\mathrm{cat}}}}P_1 e^{-D(L-1)},
\end{equation}

where $D\equiv -1/\ln \Big[{{{1}\over{2k_r}} \Big( \alpha+ k_f+k_r+f_{\mathrm{cat}} \pm \sqrt{(\alpha+k_f-k_r)^2+f_{\mathrm{cat}}(2\alpha+2(k_f+k_r)+f_{\mathrm{cat}})}\Big)}\Big]$ and $P_1$ is the monomer population fraction. Interestingly, $P(L)$ is  a superposition of two exponential functions, corresponding to the equilibrium and nonequilibrium contributions, respectively. This explains previous observations in which a single exponential distribution was not sufficient to fit $P(L)$ generated from numerical calculations \cite{microtubule_length}. As expected, Eq. 10 reduces to the equilibrium distribution if $\alpha = 0$.
Fig.3C,right shows the mean microtubule length as a function of catastrophe rate as predicted by Eq. 10 using measured rate constants \cite{microtubule_howard}, for varying levels of driving (i.e. $\alpha$, which in this case is controlled via the GTP concentration); mean lengths measured under different $f_\mathrm{cat}$ conditions in \cite{mitchison} are shown as circles. Eq. 10 predicts that, as the system is driven from equilibrium, the length distribution jumps between two distinct regimes with qualitatively different dependence on the system parameters. The jump occurs when $\alpha$ exceeds $k_r - k_f$. The suddenness of the jump is inversely proportional to $f_{\mathrm{cat}}$. In the first regime, for which $(\alpha+k_{\mathrm{f}}-k_r)/f_{\mathrm{cat}} >> 1$, Eq.10 simplifies to $<L>_{\mathrm{strong}} = (\alpha+k_{\mathrm{f}}-k_{\mathrm{r}})/f_{\mathrm{cat}}$, which is the well-known formula cited above; at physiological GTP concentrations, the predicted mean length is in excellent agreement with measured lengths \cite{mitchison} in both mitosis and interphase (circles in Fig.3C, right). Therefore, the mean length is proportional to $\alpha$ and inversely proportional to $f_{\mathrm{cat}}$ in this "strongly-driven" regime. In the second regime, for which $-(\alpha+k_{\mathrm{f}}-k_r)/f_{\mathrm{cat}} >> 1$, Eq.10 gives the length dependence of $<L>_{\mathrm{weak}}=  -\ln{[{{\alpha+k_f}\over{kr}}+{{f_{\mathrm{cat}}(\alpha+k_f)}\over{kr(\alpha+k_f-k_r)}}]}^{-1}$; in this weakly-driven regime, the mean length is only marginally sensitive to changing the driving rate or catastrophe rate (Fig.3c, right). The jump between these two regimes as $\alpha$ is increased can be made arbitrarily abrupt by decreasing the catastrophe rate. Thus, this system provides an example in which unique non-equilibrium properties (here, control of mean microtubule length) become suddenly exposed when the system is driven above a critical threshold.  
These results demonstrate the explanatory power of the probability flow equation in both recapitulating known mathematical models as well as providing new insight into their origins and complex behaviors.
\par

\section{General theorems and principles}
 Due to the generality of the probability circuit representation of nonequilibrium systems, theorems previously derived for electric circuits take on new meaning for probability flow in state space. In this section, I establish two theorems governing the energy efficiency and the reciprocal response of arbitrary systems driven away from equilibrium by an energy source. In the absence of mass sources (corresponding to grounds, See Fig.1), such systems achieve unique steady-state behavior \cite{Akle}. These theorems rely on the directional symmetry of the resistance, $R_{mn} = R_{nm}$, which holds even when the network is being driven such that the directional symmetry of the probability flow (i.e. detailed balance) is broken. In some cases, the weakly-driven (near-equilibrium) limits of these general relations correspond to known results such as Onsager's reciprocal relations. In the strongly driven regime, these relations predict qualitatively different behavior, and set hard limits on the performance capability of nonequilibrium processes in general. 

\subsection{Maximum Efficiency of Molecular Processes}
Carnot showed that the maximum energy efficiency of any heat engine is determined by the ratio between the temperatures of the heat supply and heat sink, regardless of the details of the engine \cite{carnot}. This fundamental thermodynamic relation frames our understanding of the prototypical macroscopic engine. In contrast, most molecular machines, including the proteins involved in metabolism and information processing inside cells, are chemical engines: they harness energy from a chemical fuel source to drive net reactions. Consider the circuit representation of a chemical engine, such as a molecular motor, powered by driving an equilibrium transition by an additional rate constant $\alpha_{\mathrm{d}}$. We can group all the undriven transitions into a single load resistor. The load is therefore the total resistance of all the reactions being powered "downstream" of the battery $\alpha_{\mathrm{d}}$. For example, the load could correspond to a particular cellular pattern, protein assembly, or directed movement, that is highly improbable at equilibrium; Prigogine called these "dissipative structures" because energy must be constantly consumed in order to maintain them \cite{prigogine2}. Just as in electrical circuits, the power dissipated in the load is equal to the total rate of power produced minus the power dissipated in the battery. I show here how the probability flow Eq. 5 leads to simple expressions for the energy dissipation of the battery and the load, and a general formula for the maximum efficiency of a chemical engine. 
\par
The rate of energy (or power) dissipated by the system to the environment is equal to $\sigma T$, where $\sigma$ is the entropy production rate. The fluctuation theorem provides the link between state transition probabilities and entropy production: $ {{\pi (i \rightarrow j)} / {\pi(j \rightarrow i)}} = \mathrm{e}^{- \Delta S_{i\rightarrow j}}$, where $\pi(i \rightarrow j)$ is the probability of the system transitioning from state $i$ to state $j$, and the entropy, $S$, is understood to be in units of $k_{\mathrm{B}}$ \cite{evans}. The entropy production rate is: $\sigma = \sum_{ij} I_{ij} \ln {\Bigg[ {{P_i(k_{ij}+\alpha_{ij})}\over{P_j k_{ji}}} \Bigg]}$ \cite{schnakenberg}, and can be rewritten as $\sigma=\sum_{ij} I_{ij} (\ln[P_i/P_i^*] - \ln[P_j/P_j^*]) + \sum_{ij} I_{ij} \ln[1+\alpha_{ij}/k_{ij}].$  Each term in the first summation is the current between adjacent states multiplied by the difference between a function evaluated at those states: Tellegen's Theorem dictates that such a sum equals zero at steady state \cite{tellegen}. The steady-state entropy production rate is thus:
\begin{equation}
 \sigma= \sum_{ij} I_{ij} \ln {\Bigg[1+ {{\alpha_{ij}}\over{k_{ij}}} \Bigg]}.
 \end{equation} 
Therefore, the entropy production rate is a function of just the currents of the directly driven transitions (for which $\alpha_{ij} > 0$), even though entropy is produced at all of the transitions. Eq. 11 is a useful formulation for living systems, for which the driven transitions are sparse.  \par

 The entropy production rate of the load at steady state is (See Supp. materials):
\begin{equation}
\sigma_{\mathrm{load}}=  I_{\mathrm{d}} \ln \Bigg[ {{1+({{R_{\mathrm{d}, \mathrm{Load}}}\over{R_{\mathrm{d}}+R_{\mathrm{d}, \mathrm{Load}}}}){{\alpha_{\mathrm{d}}} \over {k_{\mathrm{d}}}}}}  \Bigg],
\end{equation} 
where the load resistance, $R_{\mathrm{d},\mathrm{Load}}$, is the total resistance of the rest of the circuit if the transition d was being driven; in circuit theory the load resistance is also called the Thevenin equivalent resistance (Fig. 4a) \cite{Thevenin2}.

Near equilibrium, substituting $I_{\mathrm{d}} \approx (R_{\mathrm{d}}+R_{\mathrm{d},\mathrm{Load}})^{-1} (\alpha_{\mathrm{d}}/k_{\mathrm{d}} )$ into Eq. 12 and only keeping first order terms in $\alpha/k$ yields $\sigma_{\mathrm{Load}} ={{R_{\mathrm{d, Load}}}\over{(R_{\mathrm{d}}+R_{\mathrm{d},\mathrm{Load}})^2}}\bigg({{\alpha_{\mathrm{d}}} \over {k_{\mathrm{d}}}}\bigg)^2  \leq {{1}\over{4R_{\mathrm{d}}}} \bigg({{{\alpha_{\mathrm{d}}} \over {k_{\mathrm{d}}}}}\bigg)^2$. The maximum dissipation at the load is achieved in this inequality if the load resistance is equal to the driven resistance. This condition is analogous to the maximum power transfer theorem in electrical circuits, in which the load resistance must be equal to the internal resistance of the battery to achieve maximum power transfer to the load \cite{maxpowertransfer}. For electrical circuits, the power source is typically optimized separately from the load, allowing each to be independently optimized and combined in a modular fashion. The same principle could also be applied to the design of probability circuits. \par

The maximum steady state efficiency is the ratio of the dissipation in the load divided by the total dissipation (which is equal to the net power supplied): $\eta_{\mathrm{max}} \equiv \sigma_{\mathrm{load}}/ \sigma$. The current cancels from the numerator and denominator, yielding a maximum efficiency that is dependent only on the constitutive parameters: 
\begin{equation}
 \eta_{\mathrm{max}} = {{{ \ln \bigg[1+({{R_{\mathrm{d},\mathrm{Load}}}\over{R_{\mathrm{d}}+R_{\mathrm{d}, \mathrm{Load}}}}){{\alpha_{\mathrm{d}}} \over {k_{\mathrm{d}}}} \bigg] }} \over {{ \ln \bigg[1+{{\alpha_{\mathrm{d}}} \over {k_{{\mathrm{d}}}}} \bigg] }}}.
 \end{equation} 
This efficiency limit holds for all singly-driven systems, with the system-dependence entirely encapsulated by the single effective load resistance (Fig. 4a). Note that, just as heat engines typically perform below the Carnot efficiency, the efficiency of chemical engines can be lower than $\eta_{\mathrm{max}}$ due to system-specific waste, such as futile cycles, within the load. Eq. 13 expresses a causative relationship between efficiency (a non-additive collective property) and the constituent parameters: namely, the driven rate constant and the load resistance (which only depend on the equilibrium parameters of the system). Eq. 13 is also a tight bound, and is valid for any chemical engine regardless of the details of energy transduction. These properties distinguish Eq. 13 from previous bounds on efficiency relative to other collective properties such as fluctuations, which are typically not tight \cite{seifert_efficiency1,seifert_efficiency2}. Eq. 13 predicts that the maximum efficiency increases monotonically to unity as the system is driven farther from equilibrium; this explains previous observations that isothermal ratchets and motors are highly efficient \cite{f1atpase_oster}, and become more efficient in the strongly-driven regime \cite{efficiency, motor}. This behavior of chemical engines is reminiscent of the Carnot efficiency for heat engines, which increases if there is a high temperature differential between the heat source and heat sink \cite{carnot}. However, there are two main differences between the Carnot efficiency and the chemical engine efficiency Eq. 13. First, the chemical engine efficiency never falls below $R_{\mathrm{load}}/(R_{\mathrm{d}}+R_{\mathrm{load}})$ even when driven very close to equilibrium, meaning that systems can be efficient even if weakly driven, which is impossible for a heat engine. Second, unlike temperature differentials, the driven rate constant $\alpha_{\mathrm{d}}$ can feasibly be many orders of magnitude larger than the equilibrium rate constant $k_{\mathrm{d}}$. For example, in the cell $\alpha_{\mathrm{d}}$ usually corresponds to the rate of binding to a molecule of ATP (or GTP); biological processes are thus driven far from equilibrium by maintaining the ATP concentration above that of ADP despite the latter being about 60kJ/mol lower in energy: $\alpha_{\mathrm{d}}/k_{\mathrm{d}} \approx 10^{10.4}$ \cite{atp}. Therefore, in contrast to man-made heat engines, living systems are intrinsically capable of being highly efficient over a wide range of load resistances corresponding to diverse cellular functions (Fig. 4b). 
 
\subsection{Reciprocal Relation for Arbitrarily Driven Systems}
If the $ij$ transition in a circuit is directly driven by a voltage source $\mathcal{E}_{ij}$, then the driven rate constant can be written as a function of the current that it induces: $\alpha_{ij}[I]$. In this functional notation, the location of the induced current is given by the subscript of the function (in this case $I = I_{ij}$). In addition, nonlocal currents would be indirectly induced between other connected states in the rest of the circuit. I will denote such a nonlocal induced current, say between states $m$ and $n$, as $I_{ij \rightarrow mn} [I]$, where the location of $I$ is taken to be that of the directly driven current (in this case $I = I_{ij}$; Fig. 5). In the weakly-driven case, the indirectly induced currents are approximately linearly proportional to the driven rate constant: $I_{ij \rightarrow mn} [I]=L_{ij\rightarrow mn}\alpha_{ij}[I]/k_{ij}$, where $L_{ij\rightarrow mn}$ is the proportionality constant. Alternatively, if the driven transition were from $m$ to $n$ instead of from $i$ to $j$, there would likewise be an indirect current induced between $i$ and $j$: $I_{mn \rightarrow ij} [I]=L_{mn\rightarrow ij}\alpha_{mn}[I]/k_{mn}$.  Onsager famously showed that \cite{Onsager}:
\begin{equation}
L_{mn\rightarrow ij}=L_{ij\rightarrow mn}. 
\end{equation}
That is, the linear response to driving is the same if cause and effect are swapped. For example, in the thermoelectric effect, the electric current is induced with the same sensitivity to a temperature gradient as the heat flow is induced by an electrochemical potential \cite{Callen}. Eq. 14 is valid for multiple driven sources as well as any observables that are functions of currents because the batteries are independent (i.e. feedback is neglibile) near equilibrium, enabling superposition of multiple batteries. However, beyond the weakly-driven regime, no general reciprocal relationship between cause and effect has been established.   
In light of the mapping between nonequilibrium systems and probability potential circuits, the PFE Eq. 5 obeys Lorentz's reciprocity theorem \cite{lorentz}, which is valid for any electronic circuits in which the circuit elements are directionally symmetric. This theorem dictates that the current between $m$ and $n$ due to a fixed battery between $i$ and $j$ is equal to the current induced between $i$ and $j$ if the same fixed battery were instead placed between $m$ and $n$. For a single voltage source, $\mathcal{E}_{ij} = (R_{ij}+R_{ij,\mathrm{Load}})I_{ij}$, Eq. 5 and Lorentz reciprocity yields (See Supp. material) the general reciprocal relation:
  \begin{equation}
      I_{ij \rightarrow mn} [I]= I_{mn \rightarrow ij} \Big[{{R_{ij}+R_{ij,\mathrm{Load}}} \over {R_{mn}+R_{mn,\mathrm{Load}}}}I\Big],
  \end{equation}
which is valid arbitrarily far from equilibrium, and relates the reciprocal current responses to driving two different transitions of the system as a function of the load resistances at the two transitions (Fig. 5). For the special case in which the system is driven close to equilibrium, $P_i/P_i^* \approx 1$ and the voltage equation is approximately: $\alpha_{ij}[I]/k_{ij} \approx (R_{ij}+R_{ij,\mathrm{Load}})I_{ij}$. Substituting this relation and the definition of $L_{ij}$ into Eq. 15 recovers Onsager's near-equilibrium result Eq. 14. Intuitively, the resistance ratio in Eq. 15 describes the asymmetry in current response when the local driven transition and the nonlocal measured current are swapped. Remarkably, even if the current response becomes nonlinearly related to the voltage source far from equilibrium, the reciprocal relation between driving and response remains linear according to Eq. 15, with an invariant asymmetry ratio that is completely determined by equilibrium properties (the resistors). In any equilibrium system, the transition $ij$ with the maximum $R_{ij}+R_{ij, \mathrm{Load}}$ is the one that, if driven, induces the maximum asymmetric current response with respect to every other transition in the system, regardless of the extent of driving. This is the transition that the system is most "sensitive" to. The reciprocity relation Eq. 15 can be used to tune resistances and placement of energy sources in order to optimize desired responses to driving. Future work will determine the extent to which biological processes have evolved such that ATP and GTP binding occurs between states of maximal $R_{ij, \mathrm{Load}}$, such that the effects of driving these transitions are maximally amplified.
\par

 \par

\section{Discussion}

The Boltzmann distribution gives the relative likelihood of the states of a system at thermodynamic equilibrium regardless of the number of states or the complexity of the trajectories between states. This relation has proven immensely valuable for numerous fields including physics, chemistry, and molecular biology. Yet the vast majority of systems in nature, most famously those that comprise life, exist out of equilibrium, and their behavior has up to now been poorly and disconnectedly understood. Consequently, it has generally been difficult to answer basic questions about complex systems, such as what behaviors are permitted, how to engineer desired behaviors, and the energy required to sustain such behaviors.

Here, I derived a generalized Boltzmann distribution that is valid for all Markovian systems, in or out of equilibrium, in which the ‘standard’ Boltzmann distribution is a special case. The generalized distribution is in the form of a probability flow equation (PFE) that governs the state space "circuit" of a system, and which determines how probability currents are driven in this circuit by differences in probability potential and external driving forces. To define the potentials at each state, the probability of the state is conjugated to its equilibrium free energy, just as it is in equilibrium statistical mechanics. In addition, the currents in the circuit are conjugated to a new concept: the resistance. Zia and Schmittmann first proposed to put the currents, which are nonzero out of equilibrium, on an equal footing with the probabilities \cite{zia}. This work demonstrates that the concept of the resistance is of equal importance to that of the free energy.
\par
The resistors, which correspond to the transitions that are not directly driven, can be systematically coarse-grained, leading to an irreducible circuit whose simplicity depends on the number and placement of the driven transitions. The coarse-grained resistor in the irreducible circuit have two useful properties. First, being a function of the equilibrium parameters, they remain constant regardless of the extent of driving, and so can give insight into the invariant behavior of the system regardless of its numerical value. Second, they can be obtained without knowledge of their constituent resistors if the net current between states can be measured. This approach to top-down abstraction (Fig.2C), commonly done for electronic circuits \cite{EEbook}, is especially applicable to molecular biology, for which a minority of transitions are directly driven by an external energy supply (for example, excess ATP or GTP), thereby indirectly driving the remainder of the transitions away from equilibrium. 
\par
In addition to systems driven by energy sources, the incorporation of mass sources (states with fixed probability potentials ("grounds" in Fig.1) independent of probability flow into or out of them and nonlinear feedback (potential-dependent resistors) enable systems to achieve non-unique steady-state solutions such as multistability and oscillatory dynamics \cite{ferrell1, ferrell2}. A simple example, in which I expand the Schlogl model \cite{schlogl} by adding a tunable parallel resistor $R_{\mathrm{Load}}$ that corresponds to a catalytic pathway (Fig. 3D, left), is capable of operating as a switch transistor for binary logic (ultra-sensitivity) or one-bit erasable memory (bistability), depending on $R_{\mathrm{Load}}$. The output of this circuit is the potential drop $V_1-V_2$, which can be coupled to drive other operations in a larger circuit. The modular composition of this circuit motif, which can encode logic gates and memory, could in principle be used to engineer programmable universal computation in the same manner as electronic circuits. In tandem with the top-down approach of resistor coarse-graining, this theoretical framework may thus also enable a bottom-up approach towards scalable dynamical systems engineering.
\par
Some of the principles derived from the PFE Eq. 5, such as the maximum efficiency Eq.13 and reciprocal relation Eq.15, apply generally. Others were derived for specific classes of systems found in nature. In all cases, much of the systems' behavior could be understood with incomplete knowledge of the system's full complexity or the numerical value of system parameters. Previously unconnected known properties of weakly or irreversibly driven systems emerged as limiting cases within this framework. New properties and limits were also derived and experimental data was used to show that biological processes perform very close to these limits.  A common theme was the endowment of strongly-driven systems with properties, such as catalytic control or high energy efficiency, not accessible to weakly-driven systems, as well as the existence of all-or-none transitions between the two regimes. These findings may have implications for the necessity of establishing a large driving gradient (ATP and GTP concentrations) as a prerequisite to the subsequent optimization of biomolecular complexity. 
\par
In summary, I introduced a simple three-element circuit framework governed by the probability flow equation (PFE) that unifies nonequilibrium systems, which up to now were analyzed using a wide range of system-specific approaches. In contrast to existing approaches, the framework allows conclusions to be drawn about entire classes of systems rather than individual systems. In many cases, arbitrarily complex systems were found to obey simple rules. The ability to reason abstractly yet quantitatively opens up the possibility of modular design and evolutionary analysis based upon invariant constraints rather than individual histories.
\par

\section{Acknowledgements}
The author would like to thank Rama Ranganathan for stimulating discussions regarding this work and its implications. Andreas Doncic, Elliott Ross, Kimberly Reynolds, Luke Rice, Prashant Mishra and Arvind Murugan provided clarifying feedback. This work was supported by the Cecil and Ida Green Foundation, the Leland Fikes Foundation, and the Welch Foundation.

\pagebreak

% Following is a new environment, {scilastnote}, that's defined in the
% preamble and that allows authors to add a reference at the end of the
% list that's not signaled in the text; such references are used in
% *Science* for acknowledgments of funding, help, etc.`

\begin{figure} 
   \includegraphics[width= 14cm]{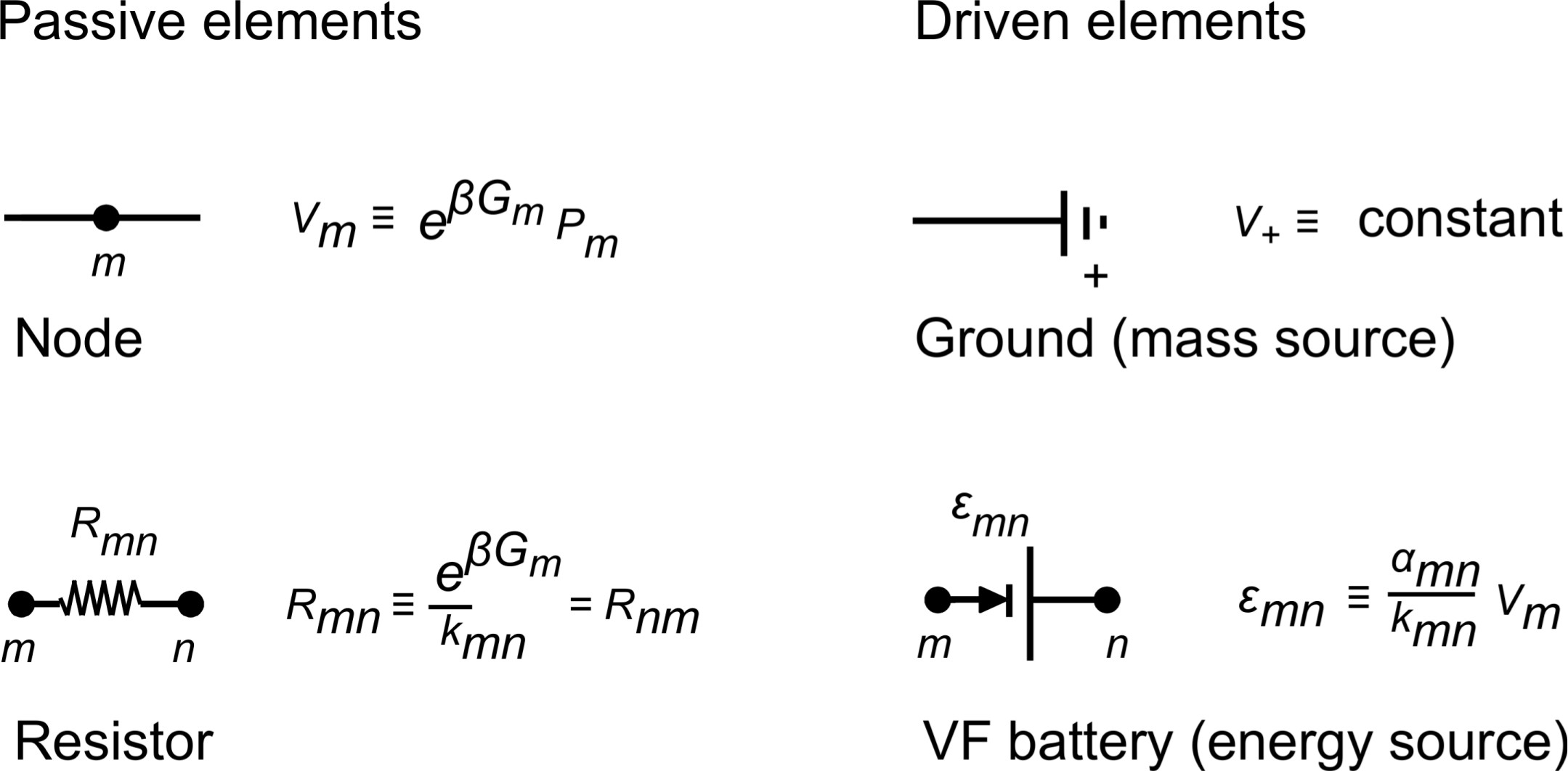}
    \caption{\textsf{\textbf{Mathematical mapping between non-equilibrium systems probability circuits state.} The potential throughout the circuit corresponds to the probability of being in different possible states. The microscopic transitions in state space correspond to elementary circuit elements; the resistors of a circuit are a function of equilibrium rate constants and free energies; and the batteries and grounds correspond to sources of energy and matter, respectively, that drive the system out of equilibrium. These circuit elements obey the probability flow equation 5 along any path in state space}}
    \label{fig1}
\end{figure}

\begin{figure} 
   \includegraphics[width= \linewidth]{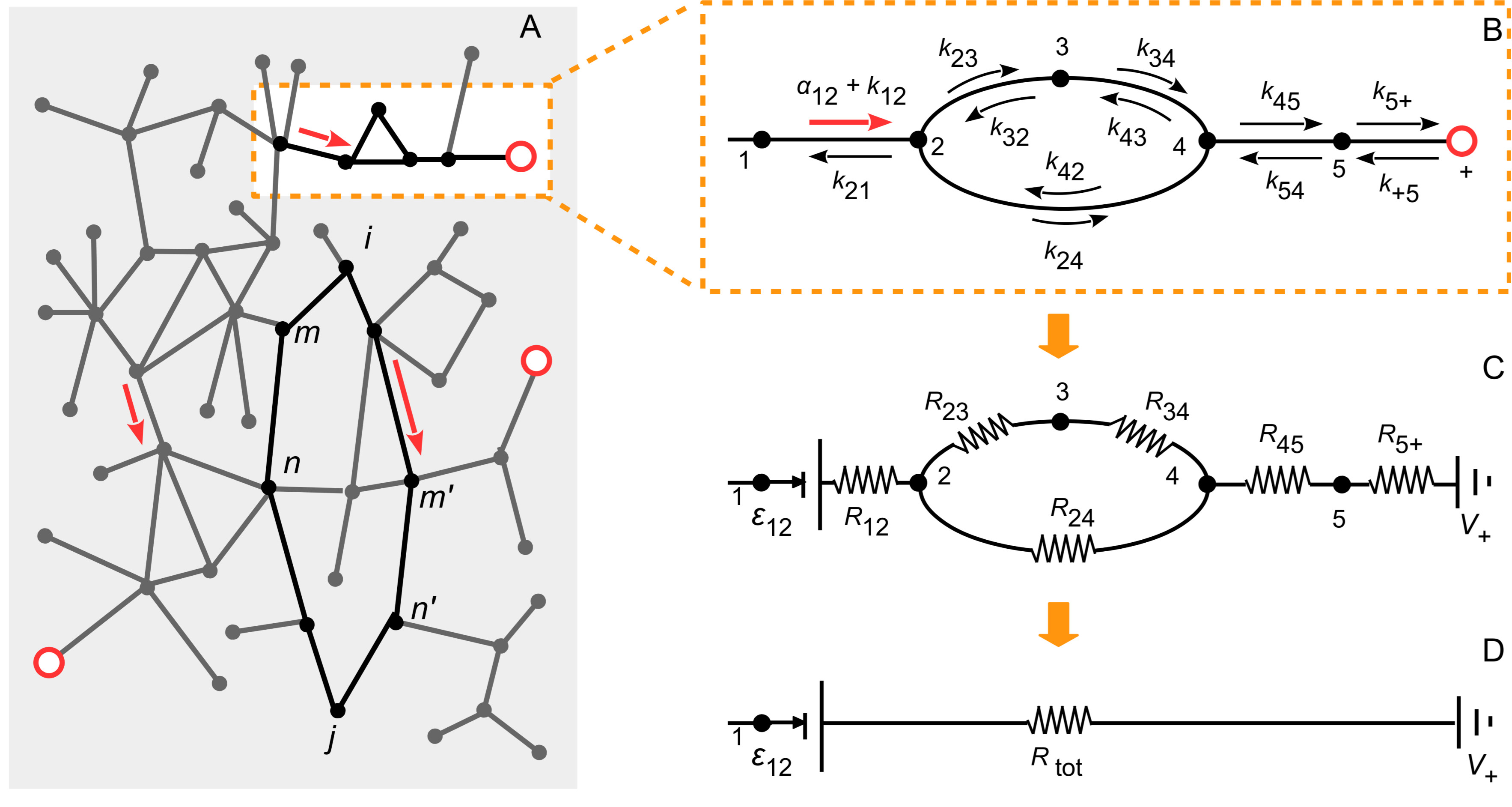}
    \caption{\textsf{\textbf{Top-down simplification of stochastic systems.} The nodes and edges represent states and transitions between states (A). All of the transitions are described by forward and reverse rate constants. At equilibrium, no net flows occur and the probability of occupying any state is given by Eq. 1. Injections of matter (red circles) or energy (red arrows) at various states and transitions, respectively, drive the system to out of equilibrium (A, B), and is governed by the PFE Eq. 5. The PFE applies to the circuit mapping (C), for which the passive components (resistors) can be simplified in a top-down manner using circuit modularization rules (D).}}
    \label{fig2}
\end{figure}

\begin{figure}
    \centerline{\includegraphics[width=\linewidth]{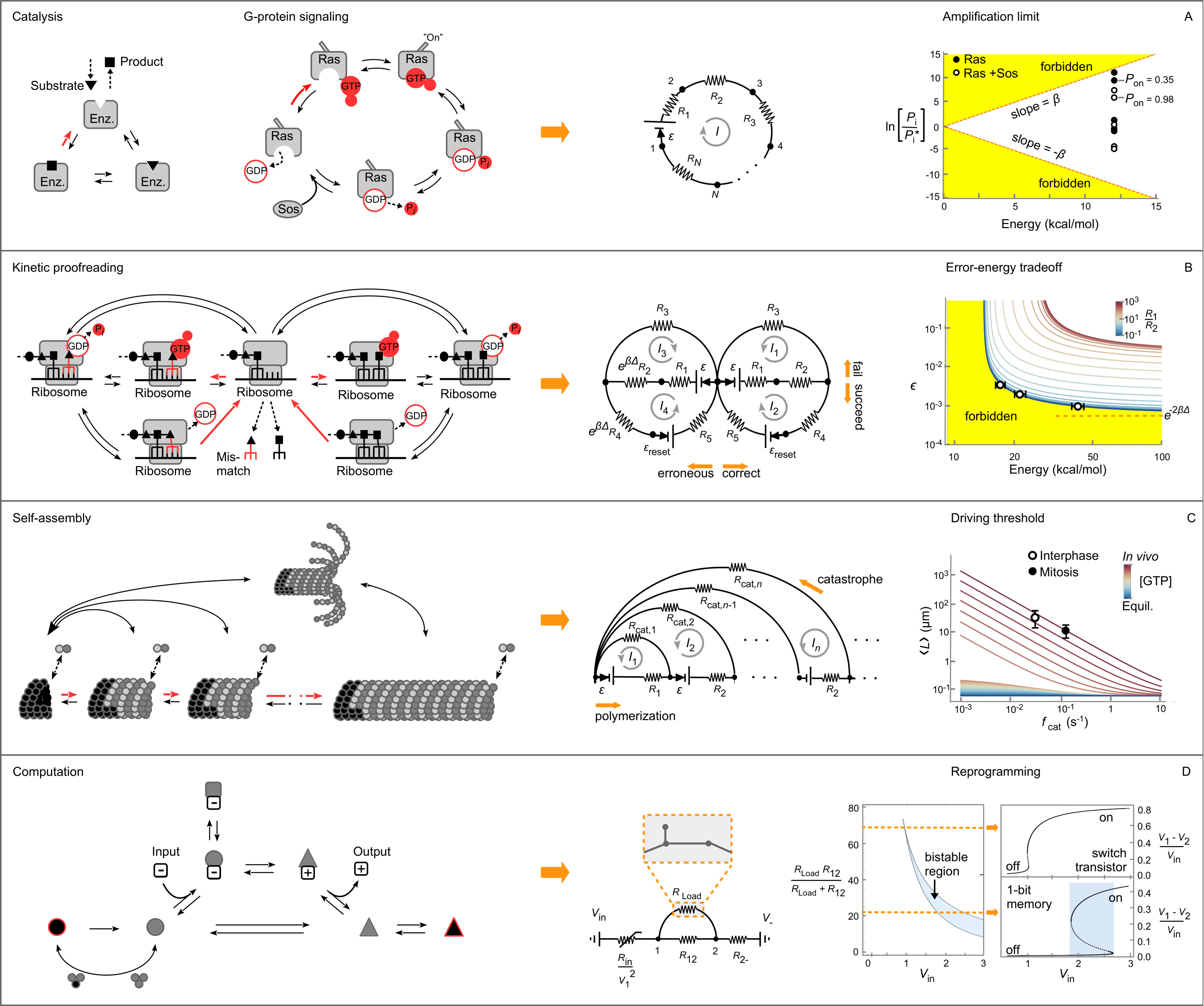}}
    \caption{\textsf{\textbf{The circuit mapping reveals hidden principles of common nonequilibrium processes.} The processes, circuit representation, and principles derived from the probability flow equation are shown from left to right. Protein catalysts and signalling molecules (A, left) belong to the broad class of systems with a cyclic state space (A, center). The extent to which desired states of such systems can be (de)amplified is bounded by the energy expenditure per cycle, as described by Eq. 8 (yellow regions are forbidden; A, right). Some states of the master signalling protein Ras approaches this limit, including the GTP-bound ("on") state, which is amplified many orders of magnitude above equilibrium levels due to net hydrolysis of GTP (A, right). The tuning of the on state probability by the production of the regulator protein Sos corresponds to changing the value of the resistors in the circuit. (B, left) Kinetic proofreading during protein translation corresponds to the driven formation of the activated complex following the correct or erroneous (energetically unfavorable by $\Delta$) codon binding which can succeed or fail, with the latter more likely for erroneous binding. Eq. 9 predicts the optimal tradeoff between accuracy and catalytic efficiency (boundary of yellow region, B, right). The optimal accuracy is achieved for falling-off rates much greater than the driven rate and is realized in the discrimination of the AAA codon over the erroneous AAU which traces the tradeoff boundary as the magnesium ion concentration is changed (open circles in B, right). (C, left) One-dimensional driven self-assembly with complete dis-assembly events (catastrophes) describes the canonical behavior of microtubules, which can be mapped to a circuit diagram with unlimited number of states (C, center). (C, right) Eq. 10 gives the mean length as a function of catastrophe frequency as well as the driving rate, which corresponds to the GTP concentration (color bar). At equilibrium (blue line), the catastrophe frequency has no influence on the mean microtubule length distribution. At physiological GTP concentration (red), the predicted frequency dependence is in excellent agreement with the measured mean microtubule length at interphase and mitosis, which differ only in the catastrophe frequency (C, right). Properties that are forbidden at equilibrium, such as the ability for a catalyst to change steady-state properties, are activated in an all-or-none fashion at a critical driving threshold (GTP concentration). The combination of energy and mass sources (i.e. batteries and constant potential nodes, respectively) with nonlinear feedback (D, left) allows the basic circuit elements to form modular components (D, center). By varying the input voltage or internal resistance such a system can be  tuned to be ultrasensitive or bistable, corresponding to a switch transistor and 1-bit memory, respectively (D, right). }}
    \label{fig3}
\end{figure}

\begin{figure}
    \centerline{\includegraphics[width=\textwidth]{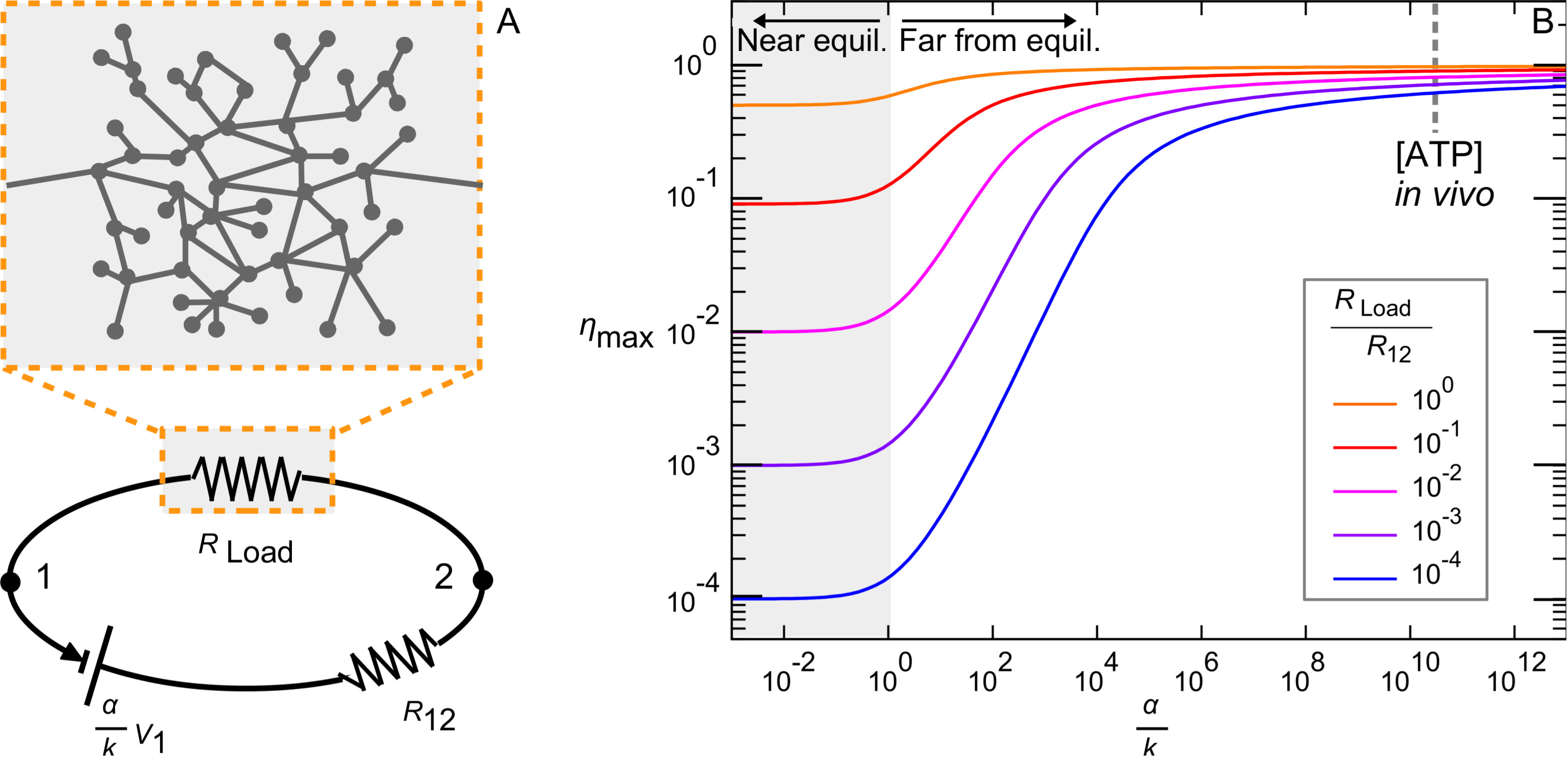}}
    \caption{\textsf{\textbf{Maximum efficiency of chemical engines.} Any equilibrium system can be driven by biasing a transition in the microstate circuit of the system's phase space, thereby breaking detailed balance. The circuit can then be divided into the driven component, which is the energy source, and the load resistance comprised of the rest of the circuit (also called Thevenin equivalent resistance) (A). Eq. 13 predicts the maximum achievable efficiency, which is system-dependent near equilibrium but tend to unity far from equilibrium, as reflected in the typical concentration levels of the energy currency in living systems, ATP (B).}}
    \label{fig4}
\end{figure}

\begin{figure}
   \includegraphics[width=\linewidth]{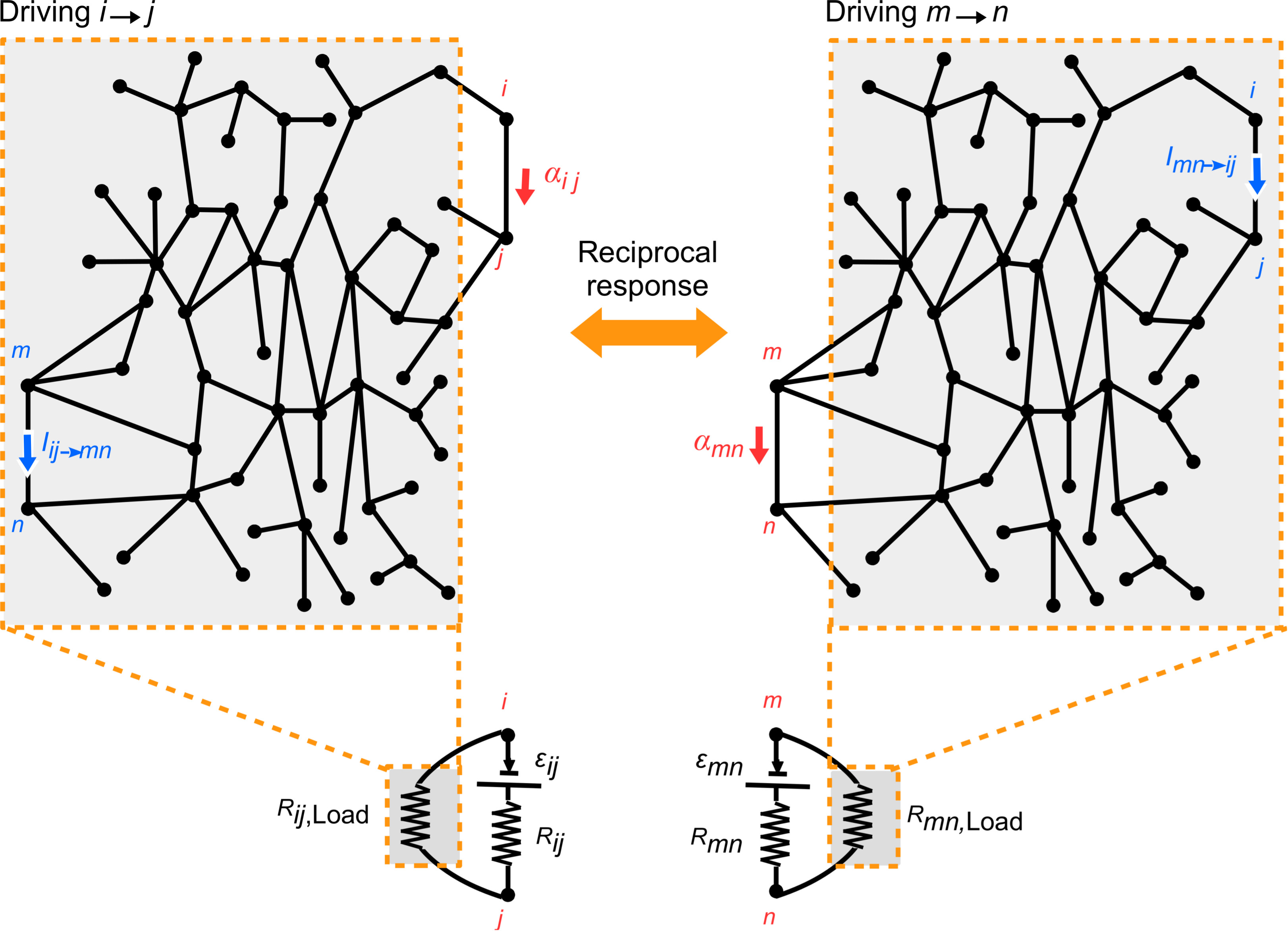}
    \caption{\textsf{\textbf{Reciprocal relations in the nonlinear regime.} If transition $ij$ in the state space of system is driven from equilibrium, there will be induced currents, for example at $mn$. If the same system were driven at $mn$ instead, then the induced current at $ij$ obeys the reciprocal relation given by Eq. 15. The current response is proportional to the relative load resistances of the driven transition and response transition. The states corresponding to the driven and response transitions are denoted in red and blue, respectively. }}
    \label{fig5}
\end{figure}

\end{document}

\title{Supplementary Information: Generalized Boltzmann distribution for systems out of equilibrium}

\author{Milo M. Lin}
\affiliation{The Green Center for Systems Biology, Department of Biophysics, and the Center for Alzheimer's and Neurodegenerative Diseases, University of Texas Southwestern Medical Center, Dallas, TX 75390 \\
Milo.Lin@UTSouthwestern.edu}

\maketitle

\title{Supplementary materials: Generalized Boltzmann distribution for systems out of equilibrium}

\section{Probability flow equation}
Starting from the definition of the current:

\begin{equation}
I_{mn} = P_m(k_{mn}+\alpha_{mn})-P_n k_{nm},
\end{equation}

multiplying both sides by ${{{\mathrm{e}^{\beta G_m}} \over {k_{mn}}}}$, and noting that the equilibrium rate constants obey the detailed balance condition: ${{k_{mn}} \over {k_{nm}}} = {e}^{\beta (G_m-G_n)}$, we obtain: 
 \begin{equation}
P_{n} \mathrm{e}^{\beta G_n}  - P_{m} \mathrm{e}^{\beta G_m} = {\alpha_{mn} \over {k_{mn}}} P_{m} \mathrm{e}^{\beta G_m} -{{{\mathrm{e}^{\beta G_m}} \over {k_{mn}}}}I_{mn} .
\end{equation} 

Along any path from state $i$ to $j$, each linkage along the path, denoted by adjacent states $m$ and $n$, satisfies this relation. Adding these equations together, all intermediate states between $i$ and $j$ cancel on the right-hand-side to give:
\begin{equation}
P_{j} \mathrm{e}^{\beta G_j}  - P_{i} \mathrm{e}^{\beta G_i} = \sum\limits_{m=i}^{n=j}\Big({\alpha_{mn} \over {k_{mn}}} P_{m} \mathrm{e}^{\beta G_m} -{{{\mathrm{e}^{\beta G_m}} \over {k_{mn}}}}I_{mn}\Big).
\end{equation} 

 The constant in front of the current $I_{mn}$ is the reciprocal of the forward rate, or "transmittance," between state $m$ and $n$ at equilibrium (times a normalization constant, the partition function). It is therefore natural to define this constant as a "resistance," $R_{mn}$, between states $m$ and $n$ (Fig. 1):
 \begin{equation}
R_{mn} \equiv {{\mathrm{e}^{\beta G_m}} \over {k_{mn}}}=R_{nm}.
\end{equation} 

Note that the second equality above follows because the equilibrium forward transmittance is equal to the equilibrium backward transmittance; hence the resistance is directionally symmetric, just like the behavior of standard resistors in electronic circuits. This property motivates the mapping of the other terms to elements of a circuit. Define the probability "potential" of a state to be:
 \begin{equation}
V_{m} \equiv P_{m} \mathrm{e}^{\beta G_m}.
\end{equation} 
Intuitively, the potential of a state is (up to normalization by the partition function) its probability divided by its equilibrium probability - a driving force for probability flows. If there is a mass source or sink in the system, which is a state whose probability (i.e. potential) remains unchanged by probability flow into or out of the state, such states correspond to "grounds." Finally, define the "battery" driving transitions from $m$ to $n$ as:
 \begin{equation}
\mathcal{E}_{mn} \equiv {\alpha_{mn} \over {k_{mn}}} \mathrm{e}^{\beta G_m}P_{m},
\end{equation} 
which is proportional to the driven rate constant divided by the equilibrium rate constant and is zero when the transition between $m$ and $n$ is not driven. $\mathcal{E}_{mn}$ is also proportional to the potential at $m$; the battery is therefore a potential-feedback battery. 
\par
Using these definitions, the potential difference between any two states $i$ and $j$ can be written in the form of Eq. 5: 
\begin{equation}
V_{j} -V_{i} = {\sum\limits_{m=i}^{n=j}(\mathcal{E}_{mn}-R_{mn}I_{mn})},
\end{equation}

where the sum is over any path from state $i$ to state $j$.

\section{Steady-state probabilities and current for single loop state space}
For $N$ states in a cycle as shown in Fig.3a(center) the drop in probability potential due to the battery is equal to the current times the total resistance:
\begin{equation}
{{\alpha}\over{k_{12}}} P_1 \mathrm{e}^{\beta G_1}=I R_{\mathrm{tot}}.
\end{equation}
The potential difference between any state $i \neq 1$ and state 1 is also given by the PFE:
\begin{equation}
P_i \mathrm{e}^{\beta G_i}-P_1 \mathrm{e}^{\beta G_1} = {{\alpha}\over{k_{12}}} P_1 \mathrm{e}^{\beta G_1}-I R_{1;i-1},
\end{equation}
where $R_{a;b} = \sum_{m=a}^{b}R_{m m+1}$ is equivalent resistance between state $a$ and $b+1$. Note that $R_{ab}$ is the resistance between connected states $a$ and $b$. 
Thus,
\begin{equation}
P_i \mathrm{e}^{\beta G_i} = ({{\alpha}\over{k_{12}}}+1) P_1 \mathrm{e}^{\beta G_1}-I R_{1;i-1}=({{\alpha}\over{k_{12}}}+1) I R_{\mathrm{tot}}{{k_{12}}\over{\alpha}}-I R_{1;i-1}.
\end{equation}
The probabilities are then:
\begin{equation}
 P_1 =I P_1^*Z R_{\mathrm{tot}} {{k_{12}}\over{\alpha}},
\end{equation}
and 
\begin{equation}
P_i =I P_i^* Z[({{\alpha}\over{k_{12}}}+1) R_{\mathrm{tot}}{{k_{12}}\over{\alpha}}-R_{1;i-1}]
\end{equation}
for $i \neq 1...$. Noting that $R_{\mathrm{tot}=R_{1;i-1}+R_{i;N}}$, $P_i$ can be expressed succinctly as:

\begin{equation}
P_i =I P_i^* Z[ R_{\mathrm{tot}}{{k_{12}}\over{\alpha}}+(1-\delta_{i1})R_{i;N}]
\end{equation}

Whether at equilibrium or not, the sum of the probabilities must be equal to one:
\begin{equation}
\sum_{i=1}^N P_i =1 = I Z[\sum_{i=2}^N P_i^* ({{\alpha}\over{k_{12}}}+1) R_{\mathrm{tot}}{{k_{12}}\over{\alpha}}-\sum_{i=2}^N P_i^* R_{1;i-1}+ P_1^* R_{\mathrm{tot}} {{k_{12}}\over{\alpha}}],
\end{equation}
where $Z$ is the equilibrium partition function. Noting that $R_{\mathrm{tot}=R_{1;i-1}+R_{i;N}}$, we can simplify this to:
\begin{equation}
1 = I Z[\sum_{i=2}^N P_i^* R_{\mathrm{tot}}{{k_{12}}\over{\alpha}}+\sum_{i=2}^N P_i^* R_{i;N}+P_1^* R_{\mathrm{tot}}{{k_{12}}\over{\alpha}}]=I Z[R_{\mathrm{tot}}{{k_{12}}\over{\alpha}}+\sum_{i=2}^N P_i^* R_{i;N}].
\end{equation}
Therefore, we obtain the stead-state current Eq.6a:
\begin{equation}
I =  {{1}\over{Z}} { { {{\alpha} \over {k_{12}}} } \over {R_{\mathrm{tot}}+{{\alpha} \over {k_{12}}} \sum\limits_{i=2}^N R_{i;N} P_i^*}}.
\end{equation}
Substituting the expression for the current into the expression for $P_i$ above, we obtain Eq. 6b:
\begin{equation}
P_{i}= P_{i}^*\Bigg[{{R_{\mathrm{tot}}+(1-\delta_{i1}){{\alpha} \over {k_{12}}}R_{i;N}} \over {R_{\mathrm{tot}}+{{\alpha} \over {k_{12}}} \sum\limits_{j=2}^N R_{j;N} P_j^*}}\Bigg]
\end{equation}

\subsection{Amplification limit}
As shown in Fig.3, without loss of generality, the single driven transition is from state 1 to 2. The current in the loop can be related to the probability of state 1 through its influence on the battery:
\begin{equation}
\epsilon=P_1 \mathrm{e}^{\beta G_1}{{\alpha}\over{k_{12}}}=I R_{\mathrm{tot}}.
\end{equation}
Therefore,
\begin{equation}
{{P_1}\over{P_1^*}}=I R_{\mathrm{tot}}Z {{k_{12}}\over{\alpha}}.
 \label{eq:P2I}
\end{equation}
Starting from the probability and current Eqs. 6a-b:
\begin{subequations}
\begin{equation}
I =  {{1}\over{Z}} { { {{\alpha} \over {k_{12}}} } \over {R_{\mathrm{tot}}+{{\alpha} \over {k_{12}}} \sum\limits_{i=2}^N R_{i;N} P_i^*}}.
\end{equation}
\begin{equation}
P_{i}= P_{i}^*\Bigg[{{R_{\mathrm{tot}}+(1-\delta_{i1}){{\alpha} \over {k_{12}}}R_{i;N}} \over {R_{\mathrm{tot}}+{{\alpha} \over {k_{12}}} \sum\limits_{j=2}^N R_{j;N} P_j^*}}\Bigg]
\end{equation}

\begin{equation}
P_{i}=P_{1} \mathrm{e}^{\beta(G_1-G_i)}\Bigg[{{R_{\mathrm{tot}}+(1-\delta_{i1}){{\alpha} \over {k_{12}}}R_{i;N}} \over {{R_{\mathrm{tot}}}}}\Bigg]
\end{equation}
\end{subequations}

Thus,
\begin{equation}
 {{{1}\over{Z}} {{\alpha} \over {k_{12}}} \over {R_{\mathrm{tot}}+{{\alpha}\over{k_{12}}}(R_{\mathrm{tot}}-R_\mathrm{12})}} < I < {{{1}\over{Z}} {{\alpha} \over {k_{12}}} \over {R_{\mathrm{tot}}}}  
 \label{eq:I_ineq}
\end{equation}

and,
\begin{equation}
{{P_i}\over{P_i^*}}={{P_1}\over{P_1^*}}\Bigg[{{R_{\mathrm{tot}}+{{\alpha} \over {k_{12}}}(R_{\mathrm{tot}}-R_\mathrm{12})} \over {{R_{\mathrm{tot}}}}}\Bigg]
\end{equation}

which implies
\begin{equation}
 {{P_1}\over{P_1^*}} < {{P_i}\over{P_i^*}}.
\end{equation}

Combining with  \eqref{eq:P2I} to obtain:
\begin{equation}
 I R_{\mathrm{tot}}Z {{k_{12}}\over{\alpha}}< {{P_i}\over{P_i^*}} = I R_{\mathrm{tot}}Z {{k_{12}}\over{\alpha}}\Bigg[{{R_{\mathrm{tot}}+{{\alpha} \over {k_{12}}}(R_{\mathrm{tot}}-R_\mathrm{12})} \over {{R_{\mathrm{tot}}}}}\Bigg]
\end{equation}
Combining with \eqref{eq:I_ineq} to obtain:
\begin{equation}
  {{1}\over {1+{{\alpha}\over{k_{12}}}\Bigg(1-{{R_\mathrm{12}}\over{R_\mathrm{total}}}\Bigg)}} < {{P_i}\over{P_i^*}} < 1+{{\alpha}\over{k_{12}}}\Bigg(1-{{R_\mathrm{12}}\over{R_\mathrm{total}}}\Bigg)
 \label{eq:P_ineq}
\end{equation}

Focusing on the right hand side, which sets the bound when $P_i > P_i^*$, we obtain after substituting for $\alpha/k_{12}$ in terms of the power $\mathcal{P}$:
\begin{equation}
   {{P_i}\over{P_i^*}} < 1+(e^{{{\mathcal{P}}\over{kTI}}}-1)\Bigg(1-{{R_\mathrm{12}}\over{R_\mathrm{total}}}\Bigg) < e^{{{\mathcal{P}/I}\over{kT}}}.
\end{equation}
Therefore:
\begin{equation}
\Big| \ln\Big({{P_i}\over{P_i^*}}\Big) \Big| < {{\mathcal{P}/I}\over{kT}}.
\end{equation}

The ras hydrolysis rate constants (except for phosphate dissociation) was found from S.E. Neal \textit{et al}, J. Biol. Chem. 263: 19718-19722 (1988). The phosphate dissociation rate of 0.102 $s^{-1}$ was found from C. Allin \textit{et al}, Proc. Natl. Acad. Sci. USA 98: 7754-7759 (2001). 
the association rate constant was chosen such that the system reached detailed balance at equilibrium ratio of [ATP]/[ADP]. 

\subsection{3-state catalytic cycle and the generalized Michaelis-Menten equation}
Starting from the steady state current Eq.6a for $N=3$:
\begin{equation}
I =  {{1}\over{Z}} { { {{\alpha} \over {k_{12}}} } \over {R_{12}+R_{23}+R_{31}+{{\alpha} \over {k_{12}}} ((R_{23}+R_{31}) P_2^*+R_{31}P_3^*)}}.
\end{equation}
For the catalyst to be useful, we assume that the rate of spontaneous conversion from product to reactant is much smaller than the other forward rates even though state 1 is more energetically favorable than state 2; hence $R_{12}$ is much bigger than the other resistors. We also note that $k_{23}/k_{32} = P_3^*/P_2^*$ and $R_{12}=\mathrm{e}^{\beta G_1}/k_{12}$. The equation simplifies to:
\begin{equation}
I =  { {\alpha} \over {Z\mathrm{e}^{\beta G_1}+\alpha ((R_{23}+R_{31})\mathrm{e}^{-\beta G_3}k_{32}/k_{23}+R_{31} \mathrm{e}^{-\beta G_3})}},
\end{equation}

where $k_{12}$ is divided out. Substituting the expressions for the resistors and noting that $k_{23} = k_f[\mathrm{S}]$, $k_{32}=k_r$,$k_{31}=k_{\mathrm{cat}}$: 
\begin{equation}
I =  { {\alpha} \over {Z\mathrm{e}^{\beta G_1}+\alpha ((\mathrm{e}^{\beta G_3}/k_r+\mathrm{e}^{\beta G_3}/k_{\mathrm{cat}})\mathrm{e}^{-\beta G_3}k_{r}/(k_{f}[\mathrm{S}])+\mathrm{e}^{\beta G_3-\beta G_3}/k_{\mathrm{cat}})}}.
\end{equation}
Multiplying top and bottom by $P_1^*=\mathrm{e}^{-\beta G_1}/Z$ and simplifying:
\begin{equation}
I =  { {[\mathrm{S}] \alpha P_1^*} \over {[\mathrm{S}]+\alpha P_1^*/k_{\mathrm{cat}}((k_{\mathrm{cat}}+k_r)/k_{f}+[\mathrm{S}])}} = { {\alpha P_1^*} \over {1+\alpha P_1^*(k_{\mathrm{M}}+[\mathrm{S}])/(k_{\mathrm{cat}}[\mathrm{S}])}},
\end{equation}

where $k_{\mathrm{M}} \equiv (k_{\mathrm{cat}}+k_r)/k_{f}$ is the Michaelis constant. We multiply the probability current $I$ by the total enzyme concentration $[\mathrm{E_o}]$ to obtain the number current per volume $v = I [\mathrm{E_o}]$, and rewrite in the form of Eq. 7:
 \begin{equation}
v ={ {k_{\mathrm{cat}}[\mathrm{E_o}][\mathrm{S}]} \over {K_{\mathrm{M}}+[\mathrm{S}]}} \Bigg({{\alpha P_1^*} \over {\alpha P_1^*+ { {k_{\mathrm{cat}}[\mathrm{S}]} \over  {K_{\mathrm{M}}+[\mathrm{S}]}}}} \Bigg) ,
\end{equation}

\section{Kinetic proofreading}

At steady state, the voltage equations taken between state 1 and the reset state in the two lower simple loops in Fig.3C are:
The voltage equation between state 1 and the reset states are:
\begin{equation}
P_1 e^{\beta G_1} - P_{\mathrm{reset}}e^{\beta G_{\mathrm{reset}}}(1+\alpha_{\mathrm{reset}}/k_{\mathrm{reset}}) = -I_1 R_3+I_2 R_4
\end{equation}

\begin{equation}
P_1 e^{\beta G_1} - P_{\mathrm{reset,error}}e^{\beta G_{\mathrm{reset}}}(1+\alpha_{\mathrm{reset}}/k_{\mathrm{reset}}) = -I_3 R_3+I_4 R_4 e^{\beta \Delta}
\end{equation}
Setting the driven reset rate much faster than the other processes ($k_{\mathrm{reset}}/\alpha_{\mathrm{reset}} \equiv \delta << 1$), making use of the definition of the voltage drop ($\epsilon \equiv \alpha/k P e^{\beta G}$), and defining without loss of generality $G_1 \equiv 0$, these simplify to: 
\begin{equation}
P_1 - \epsilon_{\mathrm{reset}}/\delta = -I_1 R_3+I_2 R_4
\end{equation}

\begin{equation}
P_1 - \epsilon_{\mathrm{reset,error}}/\delta = -I_3 R_3+I_4 R_4 e^{\beta \Delta}
\end{equation}
The two close-loop voltage equations taken around the two upper simple loops in Fig.3C are:
\begin{equation}
(I_1+I_2)(R_1+R_2) + I_1 R_3 = P_1 (\alpha/k_1)
\end{equation}
\begin{equation}
(I_3+I_4)(R_1+R_2 e^{\beta \Delta}) + I_3 R_3 = P_1 (\alpha/k_1)
\end{equation}

Where the currents are mesh currents. These four equations can be solved, retaining the lowest order in $\delta$ to give the four steady state mesh currents:
\begin{equation}
I_1= - {{P_1(R_1+R_2-{{\alpha}\over{k_1}}R_4)}\over{R_3 R_4 + (R_1+R_2)(R_3+R_4)}}
\end{equation}
\begin{equation}
I_2= {{P_1(R_1+R_2+R_3+{{\alpha}\over{k_1}}R_3)}\over{R_3 R_4 + (R_1+R_2)(R_3+R_4)}}
\end{equation}

\begin{equation}
I_3= - {{P_1(R_1+e^{\beta \Delta}(R_2-{{\alpha}\over{k_1}}R_4))}\over{R_3 R_4 e^{\beta \Delta} + (R_1+R_2 e^{\beta \Delta})(R_3+R_4 e^{\beta \Delta})}}
\end{equation}

\begin{equation}
I_4=  {{P_1(R_1+e^{\beta \Delta}R_2+R_3(1+{{\alpha}\over{k_1}}))}\over{R_3 R_4 e^{\beta \Delta} + (R_1+R_2 e^{\beta \Delta})(R_3+R_4 e^{\beta \Delta})}}
\end{equation}
As a function of these currents, the speed, rate of dissipation, and error are:
\begin{equation}
\mathrm{speed}=  I_2
\end{equation}
\begin{equation}
\mathcal{P} =  kT(I_1+I_2+I_3+I_4) \ln{[1+\alpha/k_1]}
\end{equation}

\begin{equation}
\epsilon =  I_4 /I_2 
\end{equation}

which can be combined to give:
\begin{equation}
\mathcal{P}/I_2 =  kT\Bigg(1+\epsilon + {{I_1+I_3}\over{I_2}}\Bigg) \ln{[1+\alpha/k_1]}
\end{equation}

In the biologically relevant strongly-driven regime ($\alpha/k_1 >>1$), we can further simplify to consider only the highest order in $\alpha/k_1$, giving:

\begin{equation}
\epsilon_{\mathrm{strong}} =  {{(1+{{R_2}\over{R_1}}) \mathcal{P}/I_2 }\over{(e^{-\beta \Delta}+{{R_2}\over{R_1}})\mathcal{P}/I_2 -\Big( {{2e^{-\beta \Delta}}\over{1+e^{-\beta \Delta}}}+ {{R_2}\over{R_1}}\Big)(1-e^{-2\beta \Delta})kT\ln{[1+\alpha/k_1]}}} e^{-2\beta \Delta}
\end{equation}
Note that the function ${{2x}\over{1+x}} \geq x$ for $0 \geq x \geq 1$. Since $e^{-\beta \Delta} < 1$ (the correct binding is favored over the incorrect one), this means that $\Big( {{2e^{-\beta \Delta}}\over{1+e^{-\beta \Delta}}}+ {{R_2}\over{R_1}}\Big) \geq (e^{-\beta \Delta}+{{R_2}\over{R_1}})$. Consequently,
\begin{equation}
\epsilon_{\mathrm{strong}} >  {{(1+{{R_2}\over{R_1}}) \mathcal{P}/I_2 }\over{(e^{-\beta \Delta}+{{R_2}\over{R_1}})\mathcal{P}/I_2 -( e^{-\beta \Delta}+ {{R_2}\over{R_1}})(1-e^{-2\beta \Delta})kT\ln{[1+\alpha/k_1]}}} e^{-2\beta \Delta}
\end{equation}
Again making use of the fact that $e^{-\beta \Delta} < 1$, we can replace the numerator by $(e^{-\beta \Delta}+{{R_2}\over{R_1}})$ and divide out this term from the numerator and denominator to simplify to:
\begin{equation}
\epsilon_{\mathrm{strong}} >  {{\mathcal{P}/I_2 }\over{\mathcal{P}/I_2 -(1-e^{-2\beta \Delta})kT\ln{[1+\alpha/k_1]}}} e^{-2\beta \Delta}.
\end{equation}
Because $e^{-2\beta \Delta}$ is typically much smaller than 1 (on the order of 0.0001), this further simplifies to Eq. 9 in the main text:
\begin{equation}
\epsilon_{\mathrm{strong}} >  {{\mathcal{P}/I_2 }\over{\mathcal{P}/I_2 -kT\ln{[1+\alpha/k_1]}}} e^{-2\beta \Delta}.
\end{equation}
We can see that the bound is tight, with $\epsilon$ approaching the bound for large $R_2/R_1$ as shown in Fig.3B, right.
\par
In the weakly-driven regime ($\alpha/k_1 <<1$), the expression simplifies to:
\begin{equation}
\begin{split}
\epsilon_{\mathrm{weak}} =  {{(R_1+e^{\beta \Delta}R_2+R_3)(R_3R_4+(R_1+R_2)(R_3+R_4)) }\over{(R_1+R_2+R_3) \Big( R_1(R_3+e^{\beta \Delta}R_4)+e^{\beta \Delta}(R_3R_4+R_2(R_3+e^{\beta \Delta}R_4))\Big)}} \\ +{{1}\over{kT}} {{(1-e^{\beta \Delta})R_2(R_3R_4+(R_1+R_2)(R_3+R_4)) \mathcal{P}/I_2 }\over{(R_1+R_2+R_3) \Big( R_3 R_4(1+e^{\beta \Delta})+R_1(2R_3+R_4+e^{\beta \Delta}R_4)+R_2(R_3(1+e^{\beta \Delta})+R_4(1+e^{2\beta\Delta}))\Big)}}
\end{split}
\end{equation}
Because there is an extra factor of $e^{\beta \Delta}$ in the denominators of both terms, the minimum value of $\epsilon_{\mathrm{weak}}$ is obtained in the limit that $e^{\beta \Delta}$ becomes much larger than any ratio of resistors. Therefore, keeping only the highest order in $e^{\beta \Delta}$ yields a lower bound on the efficiency which is tight in the limit of large $\Delta$:
\begin{equation}
\epsilon_{\mathrm{weak}} >  {{(1-\beta \mathcal{P}/I_2)\Big(R_3(R_1+R_2)+R_4(R_1+R_2+R_3) \Big) }\over{R_4(R_1+R_2+R_3)}} e^{-\beta \Delta},
\end{equation}
which further simplifies to:
\begin{equation}
\epsilon_{\mathrm{weak}} >  (1-\beta \mathcal{P}/I_2) e^{-\beta \Delta},
\end{equation}
which is tight if $R_4 >> R_3$. In the weakly-driven limit, $\mathcal{P} = kT(I_1+I_2+I_3+I_4)\ln{[1+\alpha/k_1]} \approx kT \alpha/k_1(I_1+I_2+I_3+I_4)$, and thus $\beta \mathcal{P}/I_2 <<1$.Consequently, $1-\beta \mathcal{P}/I_2 \approx e^{-\beta \mathcal{P}/I_2}$, and we obtain the tight error bound in the weakly-driven limit:
\begin{equation}
\epsilon_{\mathrm{weak}} >  e^{-\beta \mathcal{P}/I_2} e^{-\beta \Delta}.
\end{equation}
 The translation error and catalytic efficiency for kinetic proofreading of AAA to AAU for three different magnesium concentration conditions were obtained from Ref. 32 in the main text.
 
\section{Self-assembly}
In terms of the $n$th mesh current shown in Fig. 3c, the voltage equation taken along the path of the $n$th battery is:
\begin{equation}
P_{n+1}e^{\beta G_{n+1}}-P_{n}e^{\beta G_n}= {{\alpha}\over{k_f}}P_n e^{\beta G_n} - R_n I_n.
\end{equation}
where $R_n = e^{\beta G_n}/k_f$ and $G_n=nG$. Note that $e^{\beta G} = k_b/k_f$, where $k_f$ and $k_b$ are the equilibrium forward and backward rates, respectively. Using these definitions, we can solve for the probability of the $(n+1)$th state in terms of the previous state probability and current:
\begin{equation}
P_{n+1}=\Big(1+{{\alpha}\over{k_f}} \Big) e^{-\beta G}P_n-I_n {{e^{-\beta G}}\over{k_f}}
\end{equation}

Taking the potential difference from state $n+1$ and state 1 along the catastrophe path:
\begin{equation}
P_{1}e^{\beta G_{1}}-P_{n+1}e^{\beta G_n+1}= -R_{\mathrm{cat}, n} (I_{n}-I_{n+1}),
\end{equation}
Where $R_{\mathrm{cat}, n} = e^{\beta G_{n+1}}/f_{\mathrm{cat}} = R_{n+1} ({{k_f}\over{f_{\mathrm{cat}}}})$.
Therefore, the $(n+1)$th current is:
\begin{equation}
I_{n+1}= I_n - f_{\mathrm{cat}}P_{n+1}+ f_{\mathrm{cat}}P_{1}e^{-\beta G_{n}}
\end{equation}

In vector notation, the recursive probability and current equations become:
\begin{equation}
\begin{bmatrix}
    1       & 0 \\
    f_{\mathrm{cat}}       & 1
\end{bmatrix}
\begin{bmatrix}
    P_{n+1} \\
    I_{n+1} 
\end{bmatrix}
=
\begin{bmatrix}
    \Big(1+{{\alpha}\over{k_f}} \Big) e^{-\beta G}       & -{{e^{-\beta G}}\over{k_f}} \\
    0      & 1
\end{bmatrix}
\begin{bmatrix}
    P_{n} \\
    I_{n} 
\end{bmatrix}
+
\begin{bmatrix}
    0 \\
    {{f_{\mathrm{cat}}P_{1}}\over{e^{\beta G_{n}}}}
\end{bmatrix}
\end{equation}
Multiplying both sides by the inverse of the right-hand-side matrix, the recursion relation is:
\begin{equation}
\begin{bmatrix}
    P_{n+1} \\
    I_{n+1} 
\end{bmatrix}
=
M
\begin{bmatrix}
    P_{n} \\
    I_{n} 
\end{bmatrix}
+
\begin{bmatrix}
    0 \\
    {{f_{\mathrm{cat}}P_{1}}\over{e^{\beta G_{n}}}}
\end{bmatrix}
\end{equation}
where the transition matrix $M$ is given by:
\begin{equation}
M=
\begin{bmatrix}
    \Big(1+{{\alpha}\over{k_f}} \Big) e^{-\beta G}       & -{{e^{-\beta G}}\over{k_f}} \\
    -f_{\mathrm{cat}}\Big(1+{{\alpha}\over{k_f}} \Big) e^{-\beta G}      & f_{\mathrm{cat}}{{e^{-\beta G}}\over{k_f}}+1
\end{bmatrix}
\end{equation}

The probability and current of state $n$ in terms of those of state 1 is thus:
\begin{equation}
\begin{bmatrix}
    P_{n+1} \\
    I_{n+1} 
\end{bmatrix}
=
M^n
\begin{bmatrix}
    P_{1} \\
    I_{1} 
\end{bmatrix}
+\sum_{k=0}^{n-1}(e^{\beta G}M)^k
\begin{bmatrix}
    0 \\
    {{f_{\mathrm{cat}}P_{1}}{e^{-\beta G n}}}
\end{bmatrix}
\end{equation}
Diagonalizing $M$:
\begin{equation}
M=V
\begin{bmatrix}
    \lambda_-       & 0 \\
    0      & \lambda_+
\end{bmatrix}
V^{-1}
\end{equation}
Where the columns of $V$ are the eigenvectors of $M$ and $\lambda_-$ and $\lambda_+$ are the eigenvalues of $M$:
\begin{equation}
\lambda_{\pm} = {{e^{-\beta G}}\over{2k_f}} \Bigg( \alpha+ k_f(1+e^{\beta G})+f_{\mathrm{cat}} \pm \sqrt{(\alpha+k_f-k_fe^{\beta G})^2+f_{\mathrm{cat}}(2\alpha+2(1+e^{\beta G})k_f+f_{\mathrm{cat}})}\Bigg)
\end{equation}
Note that $\lambda_- \leq 1$ whereas $\lambda_+ \geq 1$.

The transfer matrix equation is then
\begin{equation}
\begin{bmatrix}
    P_{n+1} \\
    I_{n+1} 
\end{bmatrix}
=
V
\begin{bmatrix}
    \lambda_-^n       & 0 \\
    0      & \lambda_+^n
\end{bmatrix}
V^{-1}
\begin{bmatrix}
    P_{1} \\
    I_{1} 
\end{bmatrix}
+\sum_{k=0}^{n-1}e^{\beta k G}
V
\begin{bmatrix}
    \lambda_-^k       & 0 \\
    0      & \lambda_+^k
\end{bmatrix}
V^{-1}
\begin{bmatrix}
    0 \\
    {{f_{\mathrm{cat}}P_{1}}{e^{-\beta G n}}}
\end{bmatrix}
\end{equation}
Expanding this expression and taking the geometric sum yields $P_{n}$:
\begin{equation}
P_{n}= P_1e^{-\beta G(n-1)} {{f_{\mathrm{cat}}}\over{f_{\mathrm{cat}}-\alpha (e^{\beta G}-1)}} + A_1 \lambda_-^{n-1} + A_2 \lambda_+ ^{n-1},
\end{equation}
where the $A_1$ and $A_2$ are explicit functions of the elementary parameters. For nonzero $f_{\mathrm{cat}}$ the probability monotonically decreases for larger $n$, thus the coefficient $A_2$ must be zero. Solving this boundary condition for $I_1$ and substituting into the expression for $A_1$, we obtain the length distribution (Eq. 10 in the main text):

\begin{equation}
P_{n}= {{f_{\mathrm{cat}}}\over{f_{\mathrm{cat}}-\alpha (e^{\beta G}-1)}}P_1e^{-\beta G(n-1)}+{{\alpha(e^{\beta G}-1)}\over{\alpha (e^{\beta G}-1)-f_{\mathrm{cat}}}}P_1 e^{-D(n-1)},
\end{equation}
where $D=-\ln{\lambda_-}$.
For microtubule assembly, the physiologically relevant parameters were obtained from Ref. 38 in the main text.

\section{Extended Schlogl model}
In the extended Schlogl model, the mass/probability source and sink is represented by constant $V_{\mathrm{in}}$ and $V_{-}$, respectively. The nonlinear reaction involves the transformation of the constant supply of precursor species (filled black circle in Fig.3D, left) into the species represented by state 1 (filled gray circle). The input, $V_{\mathrm{in}}$, can be controlled by adjusting the rate of production of the precursor. The output of this system is the voltage difference $\Delta V \equiv V_1-V_2$, which can be coupled some external process to drive a reaction of interest (e.g. signalling). The input-output relation of this circuit is nonlinear, and can be tuned to qualitatively different behaviors by adjusting the resistor $R_{12}$, as shown here.
The voltage equation from "in" to 1 is:
\begin{equation}
 V_{\mathrm{in}}-V_1 = {{R_{\mathrm{in}}}\over{V_{1}^2}}I_{\mathrm{in},1},
 \end{equation} 
 where the resistance between the states, $R_{\mathrm{in}}/V_{1}^2$, is dependent on the voltage of state 1. The forward reaction is proportional to the square of species 1 population times a constant supply of the precursor given by a constant $V_{\mathrm{in}}$ (mass/probability "source" state), whereas the reverse reaction is proportional to the cube of the species 1 population because formation of a trimer complex of species 1 is required to catalyze the reverse reaction back to precursor (see Fig. 3D, left).
 The voltage equation from state 1 to 2 is:
 \begin{equation}
 V_{1}-V_2 = R_{12,\mathrm{Tot}}I_{12},
 \end{equation}
 where $R_{12,\mathrm{Tot}}=R_{\mathrm{Load}}R_{12}/(R_{\mathrm{Load}}+R_{12})$ is the result of combining the constituent resistors in parallel. This process can be used to slave any signalling reaction to its input-output function if the conversion of the "input" to the "output" signal is coupled to $R_{12}$ as shown in Fig.3D, left.
 Finally, the voltage equation from state 2 to state "-" is:
 \begin{equation}
 V_{2}-V_{-} = R_{2-}I_{2-},
 \end{equation}
 Where the voltage of state "-" stays constant because it is a mass/probability "sink" state. 
 Solving these equations for the difference between currents $I_{\mathrm{in},1}-I_{12}$, which is zero at steady state, yields the cubic equation:
 
 \begin{equation}
 I_{\mathrm{in},1}-I_{12} =0= a\Delta V^3 +b\Delta V^2 + c \Delta V +d, 
 \end{equation}
 where the coefficients are:
 \begin{equation}
 a=-{{(R_{12,\mathrm{Tot}}+R_{2-})^3}\over{R_{\mathrm{in}}R_{12,\mathrm{Tot}}^3}}. 
 \end{equation}
 \begin{equation}
 b={{V_{\mathrm{in}}R_{12,\mathrm{Tot}}^2+2V_{\mathrm{in}}R_{12,\mathrm{Tot}}R_{2-}+V_{\mathrm{in}}R_{2-}^2-3V_{-}R_{12,\mathrm{Tot}}^2-6V_{-}R_{12,\mathrm{Tot}}R_{2-}-3V_{-}R_{2-}^2}\over{R_{\mathrm{in}}R_{12,\mathrm{Tot}}^2}}.
 \end{equation}
 \begin{equation}
 c={{-R_{\mathrm{in}}+2V_{\mathrm{in}}V_{-}R_{12,\mathrm{Tot}}+2V_{\mathrm{in}}V_{-}R_{2-}-3V_{-}^2R_{12,\mathrm{Tot}}-3V_{-}^2R_{2-}}\over{R_{\mathrm{in}}R_{12,\mathrm{Tot}}}}.  
 \end{equation}
 \begin{equation}
 d={{V_{-}^2(V_{\mathrm{in}}-V_{-})}\over{R_{\mathrm{in}}}}. 
 \end{equation}
 All cubic equations with real coefficients have either one or three real roots, which correspond to the fixed points of the circuit. The former corresponds to a unique steady state, whereas the latter corresponds to a bistable state in which two of the three fixed points are stable and the other is unstable. The behavior is determined by the value of the discriminant, $\delta$, with the unique steady state if
  \begin{equation}
 \delta\equiv 18abcd -4b^3d+b^2c^2-4ac^3-27a^2d^2< 0. 
 \end{equation}
 and bistable otherwise.
 In terms of the tunable parameters of the circuit, the discriminant is:
 
 \begin{equation}
 \delta = -{{(R_{12,\mathrm{Tot}}+R_{2-})^3 \Big(4R_{\mathrm{in}}+4(R_{12,\mathrm{Tot}}+R_{2-})^2V_{\mathrm{in}}^3V_{-}-R_{\mathrm{in}}(R_{12,\mathrm{Tot}}+R_{2-})(V_{\mathrm{in}}^2+18V_{\mathrm{in}}V_{-}-27V_{-}^2)\Big)}\over{R_{\mathrm{in}}^3R_{12,\mathrm{Tot}}^6}}. 
 \end{equation}
 
 Because resistors are positive, the mono-stable condition simplifies to:
 \begin{equation}
 -4R_{\mathrm{in}}+4(R_{12,\mathrm{Tot}}+R_{2-})^2V_{\mathrm{in}}^3V_{-}-R_{\mathrm{in}}(R_{12,\mathrm{Tot}}+R_{2-})(V_{\mathrm{in}}^2+18V_{\mathrm{in}}V_{-}-27V_{-}^2) < 0
 \end{equation}
 Substituting fixed values $R_{2-}=1 s^{-1}$, $R_{\mathrm{in}}=20 s^{-1}$, and the output probability potential $V_{-}0.1$, we can plot the mono- versus bi-stable regions as a function of $R_{12,\mathrm{Tot}}$ and the input potential $V_{\mathrm{in}}$, which is shown in Fig. 3D, right.

\section{Entropy production of the load for deriving the maximum efficiency}
From Eq. 11, the total rate of entropy production is for a chemical engine with a single driven transition from state 1 to 2 (which we label 'd') is:
\begin{equation}
 \sigma= I_{\mathrm{d}} \ln {\Bigg[1+ {{\alpha_{\mathrm{d}}}\over{k_{\mathrm{d}}}} \Bigg]}.
 \end{equation} 
The rate of entropy production of the driven transition (i.e. power dissipation of the battery) is:
\begin{equation}
 \sigma_{\mathrm{d}}= I_{\mathrm{d}} \ln {\Bigg[{{(\alpha_{\mathrm{d}}+k_{\mathrm{d}})P_1}\over{P_2 k_{\mathrm{21}}}} \Bigg]}.
 \end{equation}
 Therefore, the rate of entropy production of the load is:
 \begin{equation}
 \sigma_{\mathrm{Load}}= \sigma -\sigma_{\mathrm{d}}= I_{\mathrm{d}} \ln{\Bigg[{{P_2 k_{\mathrm{21}}}\over{P_1 k_{\mathrm{d}}}}\Bigg]} .
 \end{equation} 
 
multiplying the numerator and denominator inside the logarithm by $e^{\beta G_1}$ and noting that $k_{\mathrm{d}}e^{\beta G_2}=k_{21}e^{\beta G_1}$, we obtain:
\begin{equation}
 \sigma_{\mathrm{d}}= I_{\mathrm{d}}  \ln{\Bigg[{{P_2 e^{\beta G_2}}\over{P_1 e^{\beta G_1}}}\Bigg]} .
 \end{equation}
  Taking the PFE across the cycle:
 \begin{equation}
 {{\alpha_{\mathrm{d}}}\over{k_{\mathrm{d}}}}P_1 e^{\beta G_1}-I_{\mathrm{d}}(R_{\mathrm{d}}+R_{\mathrm{Load,d}}).
 \end{equation}
 Thus,
 \begin{equation}
 I_{\mathrm{d}}= {{\alpha_{\mathrm{d}}}\over{k_{\mathrm{d}}}}(R_{\mathrm{d}}+R_{\mathrm{Load,d}})^{-1}P_1 e^{\beta G_1}=0.
 \end{equation}
 Now, by the PFE across the load resistance:
 \begin{equation}
 P_2 e^{\beta G_2}-P_1 e^{\beta G_1}=I_{\mathrm{d}}R_{\mathrm{Load, d}}.
 \end{equation}

Combining these two equations to solve for $P_2$:
 \begin{equation}
 P_2 e^{\beta G_2}= P_1 e^{\beta G_1}+{{\alpha_{\mathrm{d}}}\over{k_{\mathrm{d}}}}R_{\mathrm{d,Load}}(R_{\mathrm{d}}+R_{\mathrm{d,Load}})^{-1}P_1 e^{\beta G_1}.
 \end{equation}
 Therefore,
  \begin{equation}
 {{P_2 e^{\beta G_2}}\over{P_1 e^{\beta G_1}}}= {{R_{\mathrm{d}}+R_{\mathrm{d,Load}}(1+{{\alpha_{\mathrm{d}}}\over{k_{\mathrm{d}}}})}\over{R_{\mathrm{d}}+R_{\mathrm{d,Load}}}}.
 \end{equation}
 
Substituting into the expression for the entropy production rate of the load, we finally obtain Eq. 12 of the main text:
\begin{equation}
 \sigma_{\mathrm{Load}}= I_{\mathrm{d}} \ln{\Bigg[{{R_{\mathrm{d}}+R_{\mathrm{d,Load}}(1+{{\alpha_{\mathrm{d}}}\over{k_{\mathrm{d}}}})}\over{R_{\mathrm{d}}+R_{\mathrm{d, Load}}}}\Bigg]} .
 \end{equation}

\section{Nonlinear reciprocal relation}
Starting from an undriven circuit, consider driving the transition from $i$ to $j$ with a rate constant $\alpha_{ij}$ until the system reaches a new nonequilibrium steady state, resulting in a battery with potential drop of $\mathcal{E}_{ij}={{\alpha_{ij}}\over{k_{ij}}}P_i e^{\beta G_i}$ and a current from $i$ to $j$ $I_{ij} \equiv I$. The PFE is thus:
 \begin{equation}
 \mathcal{E}_{ij}=I (R_{ij}+R_{ij,\mathrm{Load}}).
 \end{equation}
 The battery $\mathcal{E}_{ij}$ also induces current between other states, which we denote by $m$ and $n$ without loss of generality. $I$ is monotonically related to $\mathcal{E}_{ij}$. Therefore, $I_{ij \rightarrow mn} [I]$, the long-range induced current between $m$ and $n$ is a function of the direct induced current $I$. Imagine now driving driving the transition from $m$ to $n$ instead of from $i$ to $j$. Thus, the battery will be between $m$ and $n$, with a voltage drop $\mathcal{E}_{mn}$. Crucially, let's drive the transition so that the steady-state voltage drop is the same as that of the earlier battery placed between $i$ and $j$:
  \begin{equation}
 \mathcal{E}_{mn}=\mathcal{E}_{ij}.
 \end{equation}
 As in the earlier case, there is a directly driven current between $m$ and $n$ and a long-range induced current between $i$ and $j$, denoted by $I_{mn}$ and $I_{mn \rightarrow ij}[I_{mn}]$, respectively.
 Due to the directional symmetry of the resistors and the existence of a constant voltage source at steady state, Lorentz's reciprocity theorem states that:
  \begin{equation}
      I_{ij \rightarrow mn} [I]= I_{mn \rightarrow ij} [I_{mn}].
  \end{equation}
  For each steady state corresponding to the reciprocal placement of the battery, the PFE and the equality of the voltages gives:
  \begin{equation}
 I_{mn} (R_{mn}+R_{mn,\mathrm{Load}})=\mathcal{E}_{mn}=\mathcal{E}_{ij}=I (R_{ij}+R_{ij,\mathrm{Load}}).
 \end{equation}
 Using this equation to substitute for $I_mn$ in the previous equation yields the generalized reciprocal relation Eq. 15 in the main text:
 \begin{equation}
      I_{ij \rightarrow mn} [I]= I_{mn \rightarrow ij} \Big[{{R_{ij}+R_{ij,\mathrm{Load}}} \over {R_{mn}+R_{mn,\mathrm{Load}}}}I\Big],
  \end{equation}